\documentclass[useAMS,usenatbib]{mn2e}

\usepackage{amsmath,amssymb}
\usepackage{threeparttable}
\usepackage{multirow}
\usepackage{graphicx}

\def\aap  {{A\&A}}
\def\apj    {{ApJ}}
\def\apjl   {{ApJL}}
\def\apjs   {{ApJS}}
\def\aj     {{AJ}}
\def\aaa    {{A\&A}}

\def\araa   {{ARA\&A}}

\def\mnras  {{MNRAS}}

\def\pasp   {{PASP}}

\def\procspie {{SPIE}}

\def\deg{^\circ}
\def\arcsec{$^{\prime\prime}$}

\title{The SAMI Galaxy Survey: Decomposed Stellar Kinematics of Galaxy Bulges and Disks}

\author[S.~Oh et al.]
{\parbox{\textwidth}{Sree~Oh$^{1,2}$\thanks{E-mail: sree.oh@anu.edu.au}, Matthew~Colless$^{1,2}$, Stefania Barsanti$^{3}$, Sarah Casura$^{4}$, Luca Cortese$^{2,5}$, Jesse van de Sande$^{2,6}$, Matt S. Owers$^{3,7}$, Nicholas Scott$^{2,6}$, Francesco D'Eugenio$^{8}$, Joss Bland-Hawthorn$^{2,6}$, Sarah Brough$^{9}$, Julia J. Bryant$^{2,6}$, Scott M. Croom$^{2,6}$, Caroline Foster$^{2,6}$, Brent Groves$^{2,5}$, Jon S. Lawrence$^{10}$, Samuel N. Richards$^{11}$,  Sarah M. Sweet$^{2,12,13}$} \vspace{0.4cm}\\\\
\parbox{\textwidth}{$^{1}$Research School of Astronomy and Astrophysics, Australian National University, Canberra, ACT 2611, Australia\\
$^{2}$ARC Centre of Excellence for All Sky Astrophysics in 3 Dimensions (ASTRO 3D), Australia\\
$^{3}$Department of Physics and Astronomy, Macquarie University, NSW 2109, Australia\\
$^{4}$Hamburger Sternwarte, Universit\"{a}t Hamburg, Gojenbergsweg 112, 21029 Hamburg, Germany\\
$^{5}$ICRAR, The University of Western Australia, Crawley WA 6009, Australia\\
$^{6}$Sydney Institute for Astronomy (SIfA), School of Physics, The University of Sydney, NSW 2006, Australia\\
$^{7}$Astronomy, Astrophysics and Astrophotonics Research Centre, Macquarie University, Sydney, NSW 2109, Australia\\
$^{8}$Sterrenkundig Observatorium, Universiteit Gent, Krijgslaan 281 S9, B-9000 Gent, Belgium\\
$^{9}$School of Physics, University of New South Wales, NSW 2052, Australia\\
$^{10}$Australian Astronomical Optics, Macquarie University, 105 Delhi Rd, North Ryde, NSW 2113, Australia\\
$^{11}$SOFIA Science Center, USRA, NASA Ames Research Center, Building N232, M/S 232-12, P.O. Box 1, Moffett Field, CA 94035-0001, USA\\
$^{12}$Centre for Astrophysics and Supercomputing, Swinburne University of Technology, PO Box 218, Hawthorn, VIC 3122, Australia\\
$^{13}$School of Mathematics and Physics, University of Queensland, Brisbane, QLD 4072, Australia}}

\begin{document}
\date{Accepted ---. Received ---; in original form ---}

\pagerange{\pageref{firstpage}--\pageref{lastpage}} \pubyear{2019}
\maketitle
\label{firstpage}

\begin{abstract}
We investigate the stellar kinematics of the bulge and disk components in 826 galaxies with a wide range of morphology from the Sydney-AAO Multi-object Integral-field spectroscopy (SAMI) Galaxy Survey. The spatially-resolved rotation velocity ($V$) and velocity dispersion ($\sigma$) of bulge and disk components have been simultaneously estimated using the penalized pixel fitting (pPXF) method with photometrically defined weights for the two components. We introduce a new subroutine of pPXF for dealing with degeneracy in the solutions. We show that the $V$ and $\sigma$ distributions in each galaxy can be reconstructed using the kinematics and weights of the bulge and disk components. The combination of two distinct components provides a consistent description of the major kinematic features of galaxies over a wide range of morphological types. We present Tully-Fisher and Faber-Jackson relations showing that the galaxy stellar mass scales with both $V$ and $\sigma$ for both components of all galaxy types. We find a tight Faber-Jackson relation even for the disk component. We show that the bulge and disk components are kinematically distinct: (1)~the two components show scaling relations with similar slopes, but different intercepts; (2)~the spin parameter $\lambda_{R}$ indicates bulges are pressure-dominated systems and disks are supported by rotation; (3)~the bulge and disk components have, respectively, low and high values in intrinsic ellipticity. Our findings suggest that the relative contributions of the two components explain, at least to first order, the complex kinematic behaviour of galaxies.
\end{abstract}


\section{Introduction}
\label{sec:intro}
Integral field spectroscopy (IFS) enables precise measurements of the angular momentum of galaxies. Recent studies from IFS surveys have reported that galaxies show a wide range of kinematic properties that are known to be related to other observable quantities such as mass, morphology, population, and environment (e.g.\ Cappellari et~al.\ 2007; Emsellem et~al.\ 2007, 2011; Krajnovi$\acute{\rm c}$ et~al.\ 2013; Brough et~al.\ 2017; van de Sande et~al.\ 2017b; Graham et~al.\ 2018, 2019; Falc{\'o}n-Barroso et al.\ 2019). As is well established by the parametric fitting of surface brightness, many galaxies can be characterised by two distinct components, classically termed bulges and disks. They are also expected to have different kinematic signatures, reflecting the complex distribution of stellar orbits and angular momentum in galaxies.

The stellar kinematic properties of galaxies (rotation velocity and velocity dispersion) scale with stellar mass, which represents the proportionality of the stellar and dynamical masses. This is reflected in the Faber-Jackson relation between luminosity and velocity dispersion (Faber \& Jackson 1976) and the Tully-Fisher relation between luminosity and rotation velocity (Tully \& Fisher 1977). However, both scaling relations are normally considered to apply to specific types of galaxies: the Faber-Jackson relation applies to early-type (pressure-supported) galaxies, while the Tully-Fisher relation applies to late-type (rotation-supported) galaxies. Combining measurements of the rotation velocity and velocity dispersion for each galaxy is one promising approach to a unified scaling relation addressing the kinematics of all types of galaxies (e.g.\ Weiner et~al.\ 2006; Kassin et~al.\ 2007; Cortese et~al.\ 2014; Simons et~al.\ 2015; Straatman et~al.\ 2017; Aquino-Ort$\acute{\rm i}$z et~al.\ 2018; Barat et~al.\ 2019). Another approach to encompassing all types of galaxies in scaling relations is decomposing the kinematics of (pressure-supported) bulge and (rotation-supported) disk components. Early-type galaxies are bulge-dominated, and so the velocity dispersion of the bulge component drives the Faber-Jackson relation; in contrast, late-type galaxies are disk-dominated, and so the rotation velocity of the disk component drives the Tully-Fisher relation. 
 
 Photometric structure measurements can be combined with integral field spectroscopy (IFS) to spectroscopically decompose bulge and disk components and explore their individual kinematic properties and stellar populations, as well as the ways the two components determine overall galaxy properties such as the distribution of angular momentum. Such spectroscopic decomposition was employed with long-slit observations to estimate the age and metallicity of 21 lenticular galaxies in the Virgo cluster (Johnston et~al.\ 2014). Johnston et~al.\ (2017) applied a similar method to the Mapping Nearby Galaxies at Apache Point Observatory (MaNGA; Bundy et~al.\ 2015) IFS data, decomposing the two components using image slices at each wavelength. They then constructed bulge and disk spectra using the flux weights at each wavelength and presented the stellar populations of two galaxies measured with the Lick system (Burstein et~al.\ 1984; Faber et~al.\ 1985). Catal\'an-Torrecilla et~al. (2017) and M\'endez-Abreu et~al.\ (2017, 2019) also applied a spectro-photometric decomposition to three galaxies from the Calar Alto Legacy Integral Field Area survey (CALIFA; S\'anchez et~al.\ 2012). Tabor et~al.\ (2017) introduced simultaneous spectral fitting of the two components using the penalized pixel fitting code (pPXF; Cappellari \& Emsellem 2004; Cappellari 2017) and applied this to three lenticular galaxies from the CALIFA survey and subsequently to 302 early-type galaxies from the MaNGA survey; the population and kinematics of the two components have been presented in Tabor et~al.\ (2019). Coccato et~al.\ (2011, 2018) also performed full spectral fitting to decompose NGC~5719 and NGC~3521 using the pPXF and the maximum penalized likelihood method (Gebhardt et~al.\ 2000; Fabricius et~al.\ 2014).

Although there have been several methods developed for spectroscopic decomposition, they have so far mostly been applied to a handful of galaxies having a specific type (e.g.\ lenticulars), with a focus on the stellar populations of two components. Because lenticular galaxies are well described by two components without introducing complex morphological features (e.g.\ bars, spiral arms, and rings), they serve as ideal testbeds for spectroscopic decomposition. Tabor et~al.\ (2019) is the only study to date that has applied spectroscopic decomposition to a large sample and also addressed the kinematics of the two components. They measured the stellar angular momentum $j_*$ and found that bulge and disk components display distinct rotation properties. However, Tabor et~al.\ (2019) only studied early-type galaxies, leaving open a comprehensive view of the kinematics of the bulge and disk components for galaxies of all types. 

In this paper we investigate the kinematics of bulge and disk components for galaxies with a wide range of morphological types using data from the SAMI Galaxy Survey (Croom et~al.\ 2012; Bryant et~al.\ 2015; Green et~al.\ 2018; Scott et~al.\ 2018). We explore the Tully-Fisher and Faber-Jackson relations for the two components and discuss the reconstruction of the overall galaxy kinematics using the decomposed kinematics of the bulge and disk components. 

The paper is organised as follows. Section~\ref{sec:sami} introduces the SAMI survey data. Section~\ref{sec:decomp} describes the spectroscopic decomposition of the bulge and disk kinematics using pPXF with component weights from photometric fits and physically-motivated model constraints (including a new pPXF subroutine to deal with degeneracy in the solutions). In Section~\ref{sec:app}, we describe the application of this method to the SAMI data and sample. We present results on the stellar kinematics of the bulge and disk components, their scaling relations, and their spin parameters in Section~\ref{sec:res}. In Section~\ref{sec:dis}, we discuss what determines the kinematics of galaxies and summarise our conclusions in Section~\ref{sec:con}. Throughout the paper, we assume a standard $\Lambda$CDM cosmology with $\Omega_m = 0.3$, $\Omega_{\Lambda} = 0.7$, and $H_0 = 70$\,km\,s$^{-1}$~Mpc$^{-1}$.
 
\section{The SAMI Galaxy Survey}
\label{sec:sami}
SAMI (Croom et~al.\ 2012) is a multi-object fibre-integral-field system feeding the AAOmega spectrograph (Sharp et~al.\ 2006) on the 3.9-metre Anglo-Australian Telescope. The 13 15\arcsec-diameter hexabundles, each composed of 61 1.6\arcsec-diameter optical fibres, patrol a 1$\deg$-diameter field of view (Bland-Hawthorn et~al.\ 2011; Bryant et~al.\ 2011, 2014). The SAMI survey uses AAOmega's 581V and 1000R gratings in the blue (3750--5750\,\AA) and red (6300--7400\,\AA) arms of the spectrograph, giving spectral resolutions of $R$=1808 and $R$=4304 respectively (van de Sande et~al.\ 2017b).

The SAMI Galaxy Survey (Bryant et~al.\ 2015) includes 2964 unique galaxies at redshifts $0.04 < z < 0.095$; 2083 are from the Galaxy And Mass Assembly (GAMA) G09, G12, and G15 regions (Driver et~al.\ 2011), while 881 lie within the virial radius of eight rich clusters (Owers et~al.\ 2017). A series of volume-limited samples with increasing stellar mass limits at increasing redshift has been used for the SAMI-GAMA sample (Bryant et~al.\ 2012, Figure~4). The stellar masses ($M_*/M_{\odot}$) have been derived from the $i$-band magnitudes and $g - i$ colours (Taylor et~al.\ 2011). The SAMI-cluster sample uses two mass limits of $\log M_*/M_{\odot} \geq 9.5$ and $\log M_*/M_{\odot} \geq 10$ for galaxies in the clusters at $z \leq 0.045$ and $0.045 < z \leq 0.06$ respectively. SAMI typically covers one to two effective radii of the sample galaxies; more than 75\% of the sample galaxies are available to measure kinematics out to one effective radius (See Figure 1 of van de Sande et~al.\ 2017a).

The data reduction pipeline is described in Sharp et~al.\ (2015) and Allen et~al.\ (2015). The pipeline has been subsequently updated, and the current data reduction and data quality are presented in Scott et~al.\ (2018). Twelve team members have visually determined the galaxy morphology of the SAMI galaxies following the scheme in Kelvin et~al.\ (2014) and using the colour-composite images from the Sloan Digital Sky Survey (SDSS; York et~al.\ 2000) Data Release 9 (Ahn et~al.\ 2012) and the VST ATLAS survey (Shanks et~al.\ 2013), respectively, for the SAMI-GAMA and SAMI-cluster samples. The galaxies are classified into four main groups: ellipticals (E), lenticulars (S0), early spirals (S$_{\rm E}$), and late spirals (S$_{\rm L}$); see Cortese et~al.\ (2016) for more details on the SAMI visual morphology classifications.

\begin{figure}
\centering
\includegraphics[width=\columnwidth]{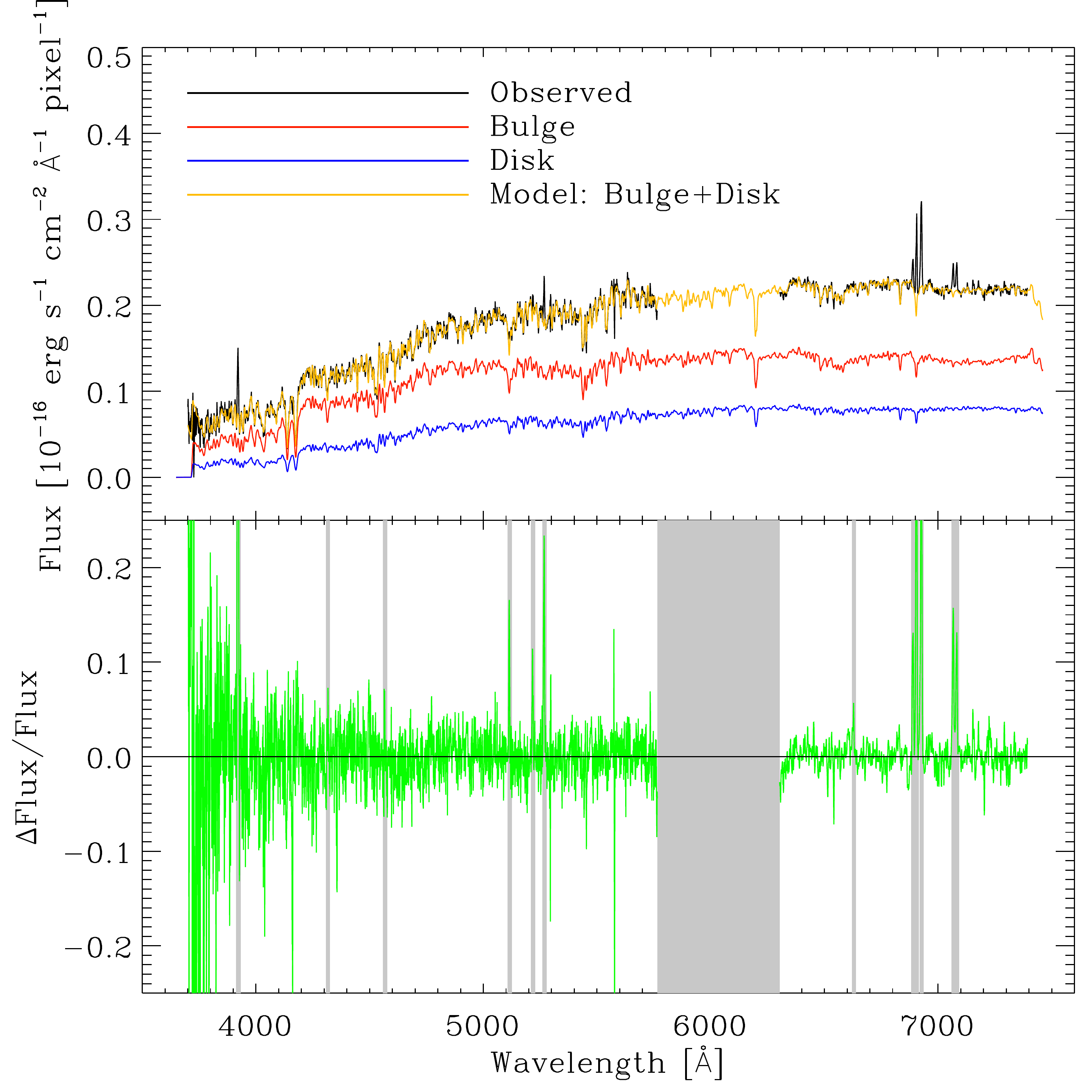}
\caption{An example of spectroscopic decomposition using pPXF. The upper panel shows the observed spectrum (black), the decomposed bulge (red) and disk (blue) spectra, and the reconstructed total spectrum model (orange). The lower panel shows the relative residuals of the fit between the observed and model spectra, which become larger at lower wavelengths due to decreasing signal-to-noise at the blue end of the spectrum. The grey shaded area indicate the initial masking of emission lines and the gap between the blue and red spectra. }
\label{ppxf}
\end{figure}   

\begin{figure*}
\centering
\includegraphics[width=\textwidth]{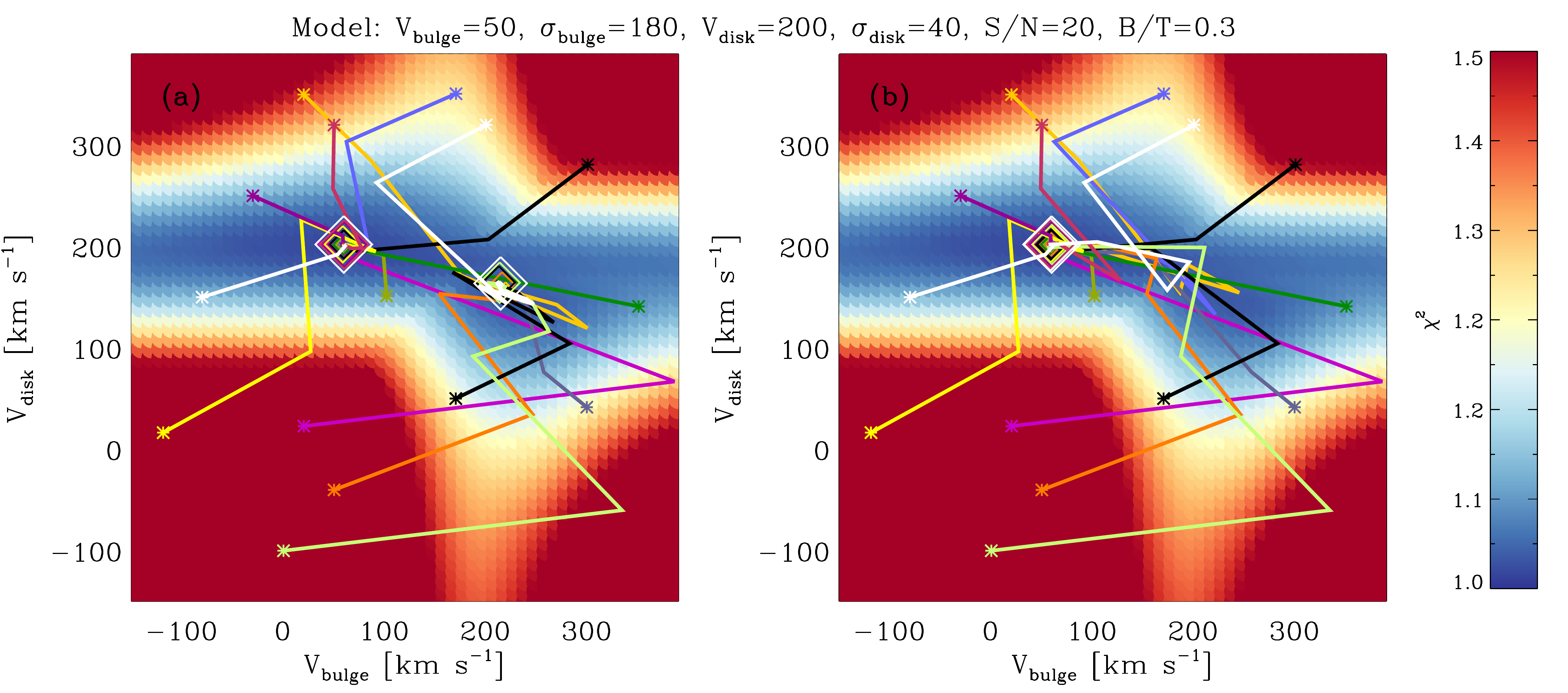}
\caption{The pPXF solutions for various initial parameter estimates (a)~without and (b)~with the swapping method. The true kinematic solution for the simulated spectrum is shown in the top label. The background has been coloured by the reduced $\chi^2$ from the pPXF runs with the fixed velocities. We selected 15 initial guesses (asterisks) and followed the trajectory to the final solutions (diamonds). When the swapping routine is turned off, pPXF sometimes generates the solution based on the local minimum of the $\chi^2$ where $|V_{\rm bulge}| > |V_{\rm disk}|$. The swapping routine forces pPXF to find the global minimum in $\chi^2$, and therefore the true solution.}
\label{swap}
\end{figure*}
     
\section{Decomposing bulge and disk kinematics}
\label{sec:decomp}
We decompose the rotation velocities and velocity dispersions of galaxies into their spatially resolved disk and bulge components. This involves a series of steps, summarised here, and described in detail in the following subsections. The key steps are setting photometric constraints, performing spectroscopic decomposition using the pPXF method with a new subroutine for physically-motivated solutions, and testing the method.

\subsection{Photometric bulge-disk decomposition}
\label{sec:phot}
Photometric bulge-disk decomposition has been performed using the $r$-band images from the Kilo-Degree Survey (KiDS; de Jong et~al.\ 2017) for the SAMI-GAMA sample (S.~Casura et~al., in preparation) and from the Sloan Digital Sky Survey (SDSS) for the SAMI-cluster sample (S.~Barsanti et~al., in preparation). The analysis employed ProFit, a routine for Bayesian two-dimensional galaxy profile modelling (Robotham et~al.\ 2017). Free S\'ersic and exponential profiles have been used for bulge and disk components respectively. The effective radius ($R_{\rm e}$), ellipticity ($\epsilon$), and position angle ($PA$) of galaxies are determined based on a single-component S\'ersic fit using the same photometric images. The two-dimensional flux models from the photometric decomposition are convolved with a kernel to match the point spread function (PSF) of the SAMI reconstructed images and rebinned to have the same pixel scale as the SAMI cubes (0.5\arcsec). Finally, we estimated the bulge-to-total light ratio for each SAMI spaxel as the ratio between the bulge model flux and the total flux, which is used as a relative flux constraint in the spectral bulge-disk decomposition in Section~\ref{sec:ppxf}. We take the overall luminosity fraction of the bulge component (B/T) measured from the entire light profile as a proxy for galaxy type. The luminosity fraction of the bulge measured within 1\,$R_{\rm e}$ (B/T$_{\rm e}$) is used to reconstruct the observed kinematics in Section~\ref{reconstruction}. 

\subsection{Spectral bulge-disk decomposition}
\label{sec:ppxf}
We use the pPXF method---a routine for full spectral fitting---to extract the rotation velocity ($V$) and velocity dispersion ($\sigma$) of both the bulge and disk components. The {\sc fraction} keyword of pPXF allows simultaneous estimation of two distinct kinematic components (such as a bulge and a disk) with a constraint on the relative weights of the two components 
\begin{equation}
\label{eq:fb1}
f_{\rm bulge} = \frac{\sum w_{\rm bulge}}{\sum w_{\rm bulge} + \sum w_{\rm disk}}, 
\end{equation}
where $w_{\rm bulge}$ and $w_{\rm disk}$ are the sum of the weights for the templates used for each component. The linear least-squares sub-problem with an exact linear equality constraint (Equation~\ref{eq:fb1}) can be simplified by adding the following extra equation requiring only minimal changes to the pPXF algorithm:
\begin{equation}
\label{eq:fb2}
(f_{\rm bulge}-1) \sum w_{\rm bulge} + f_{\rm bulge}\sum w_{\rm disk} \leq \Delta,
\end{equation}
where $\Delta$ regulates the precision required to satisfy Equation~\ref{eq:fb2} and is set to a small number (e.g.\ $10^{-9}$). Both the stellar template spectra and the galaxy spectrum have been normalised to have a mean flux of order unity to satisfy the equality constraint to sufficient numerical accuracy. Figure~\ref{ppxf} presents an example of the spectral decomposition of bulge and disk components using pPXF.
The details for the feature of {\sc{fraction}} keyword are well introduced in Cappellari (2017) and Tabor et~al.\ (2017). The photometrically-defined weights of the two components have been used as the constraint in the spectroscopic decomposition (Section~\ref{sec:phot}).

The pPXF method sometimes yields solutions where bulge and disk kinematics appear to be swapped with each other, a known issue of simultaneously fitting two components. The presence of multiple minima in the $\chi^2$ landscape sometimes traps the minimisation algorithm in a local minimum (i.e.\ a `swapped' solution). The near-degeneracy of the $\chi^2$ for this model and the corresponding features of pPXF solutions are also described in Figure~4 of Tabor et~al.\ (2017). These local minimum solutions are typically unphysical, in the sense that they imply the bulge rotation velocity is larger than the disk rotation velocity, $|V_{\rm bulge}| > |V_{\rm disk}|$, and the bulge velocity dispersion is smaller than the disk velocity dispersion, $\sigma_{\rm bulge}  < \sigma_{\rm disk}$. This suggests the need for physically-motivated constraints to resolve this near-degeneracy in the models. 

Tabor et~al. (2017, 2019) introduced a method performing decomposition 25 times for each spectrum across a 5 by 5 grid of bulge and disk velocities ranging from -350 to 350 \,km\,s$^{-1}$ to find the global minimum of the $\chi^2$. The method is expected to be efficient in most circumstances, but we found a few exceptions. First, the grid method is not effective to differentiate the global and local minima when velocities of two components are close to each other, and two minima may even lie within the same bin on a grid with a size of 140\,km\,s$^{-1}$. Second, two minima have the same $\chi^2$ values in theory when two components have the same weights of 0.5. Considering the moderate S/N of the current IFS data, the global minimum might not indicate the true kinematic solution, especially when the weights of two components are comparable to each other (i.e.\ B/T\,$\sim0.5$).

We therefore introduce a swapping routine to overcome the degeneracy resulting from two similar local $\chi^2$ minima. This routine is activated when 
\begin{equation}
\sigma_{\rm bulge} + \sigma_{\rm error} < \sigma_{\rm disk},
\label{swapcondition}
\end{equation}
where $\sigma_{\rm error}$ is the error in $\sigma$ from a single-component pPXF fit. The swapping routine uses the bounded-variable least squares (BVLS; Lawson \& Hanson 1974) algorithm, which limits the solution within the boundary condition. First, the $V$ and $\sigma$ of the two components are swapped before starting the main BVLS loop. Then the boundary condition, which is active only within the routine, is set as follows:
\begin{equation}
\left.
\begin{aligned}
|V_{\rm bulge}| <|V_{\rm bulge,old}| \\ 
 \sigma_{\rm bulge}>\sigma_{\rm bulge,old} \\
 \sigma_{\rm disk} < \sigma_{\rm disk,old} \\ \end{aligned} 
 \right\}
\label{swapupdate}
\end{equation}
where $V_{\rm bulge,old}$, $\sigma_{\rm bulge,old}$, and $\sigma_{\rm disk,old}$ are the solutions for bulge velocity, bulge velocity dispersion, and disk velocity dispersion, respectively, before the swap. It is more difficult to estimate the kinematics for a component with low signal-to-noise (S/N), so the $V$ and $\sigma$ of the component with higher S/N are initially set to be bounded variables for solving within the boundary condition (Equation~\ref{swapupdate}) while the $V$ and $\sigma$ of the other component are fixed. The S/N of the two components are estimated as B/T\,$\times$\,S/N and (1-B/T)\,$\times$\,S/N for bulge and disk components respectively. The previous steps make this a simple BVLS problem with a boundary condition and fixed and bounded variables. The BVLS main loop returns the bounded solution to the main solver of pPXF for further progress. 

If the routine is called 50 times, then pPXF stops calling the routine even if the solution still meets the swapping condition given in equation~\ref{swapcondition} and generates the solution without using the routine. We randomly selected 150 SAMI galaxies which include 58997 spectra with S/N greater than 3\,\AA$^{-1}$ and found that the swapping routine has been called 50 times on 360 spectra ($\sim 0.6\%$). To reduce unnecessary iterations, the routine is also not activated when $\sigma_{\rm bulge}$ or $\sigma_{\rm disk}$ is greater than 600\,km\,s$^{-1}$, which happens on 7293 out of 58997 spectra from the 150 randomly selected SAMI galaxies. The swapping routine does not apply a hard constraint on the kinematics, unlike the {\sc{bound}} keyword of pPXF, but re-initialises the input guesses by swapping the kinematics of the two components.

Figure~\ref{swap} shows an example $\chi^2$ distribution for a model spectrum when fitting two components (see also Figure~4 of Tabor et~al.\ 2017). We ran pPXF multiple times to extract the reduced $\chi^2$ for each grid point. Then we ran pPXF for the same model allowing a free estimation of kinematics with various combinations of input velocities (asterisks); the true velocity dispersions were used as initial guesses for simplicity. We present the trajectory to the final solution (diamonds) to test the difference between turning the swapping routine on and off. The solution from pPXF is sensitive to initial guesses without the swapping routine (left panel); pPXF sometimes returns a solution based on the local $\chi^2$ minimum where $|V_{\rm bulge}| > |V_{\rm disk}|$ and $\sigma_{\rm bulge}  < \sigma_{\rm disk}$ (i.e.\ a `swapped solution'). On the other hand, when the swapping routine is turned on (right panel), pPXF is prevented from falling into the local minimum and fits the correct solution regardless of the initial guess for the rotation velocity. 

\begin{figure}
\centering
\includegraphics[width=\columnwidth]{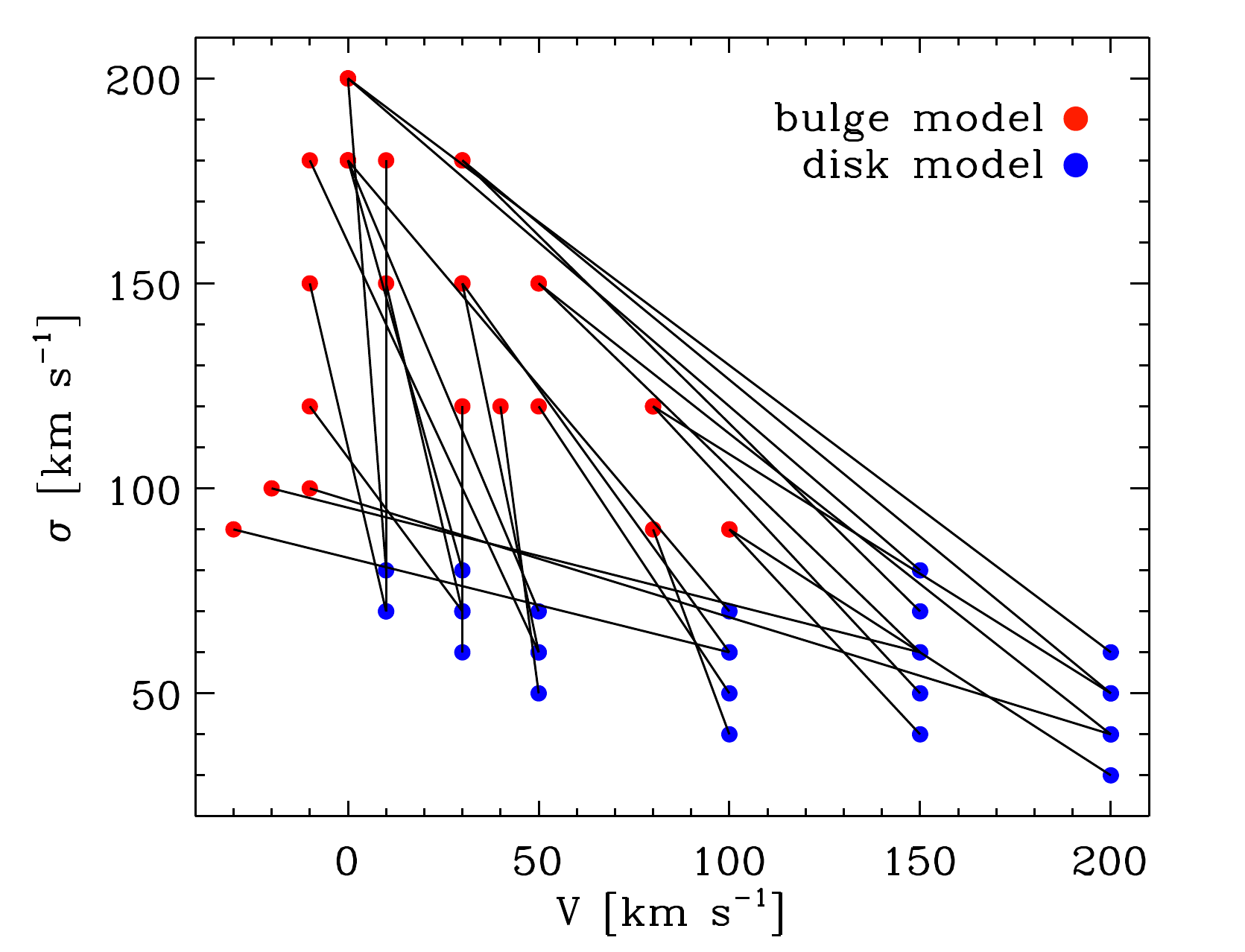}
\caption{The 28 sets of input parameters for the bulge (red) and disk (blue) kinematic models used to generate mock spectra; the paired bulge and disk parameters are linked by the solid lines.}
\label{model}
\end{figure}    

\begin{figure}
\centering
\includegraphics[width=\columnwidth]{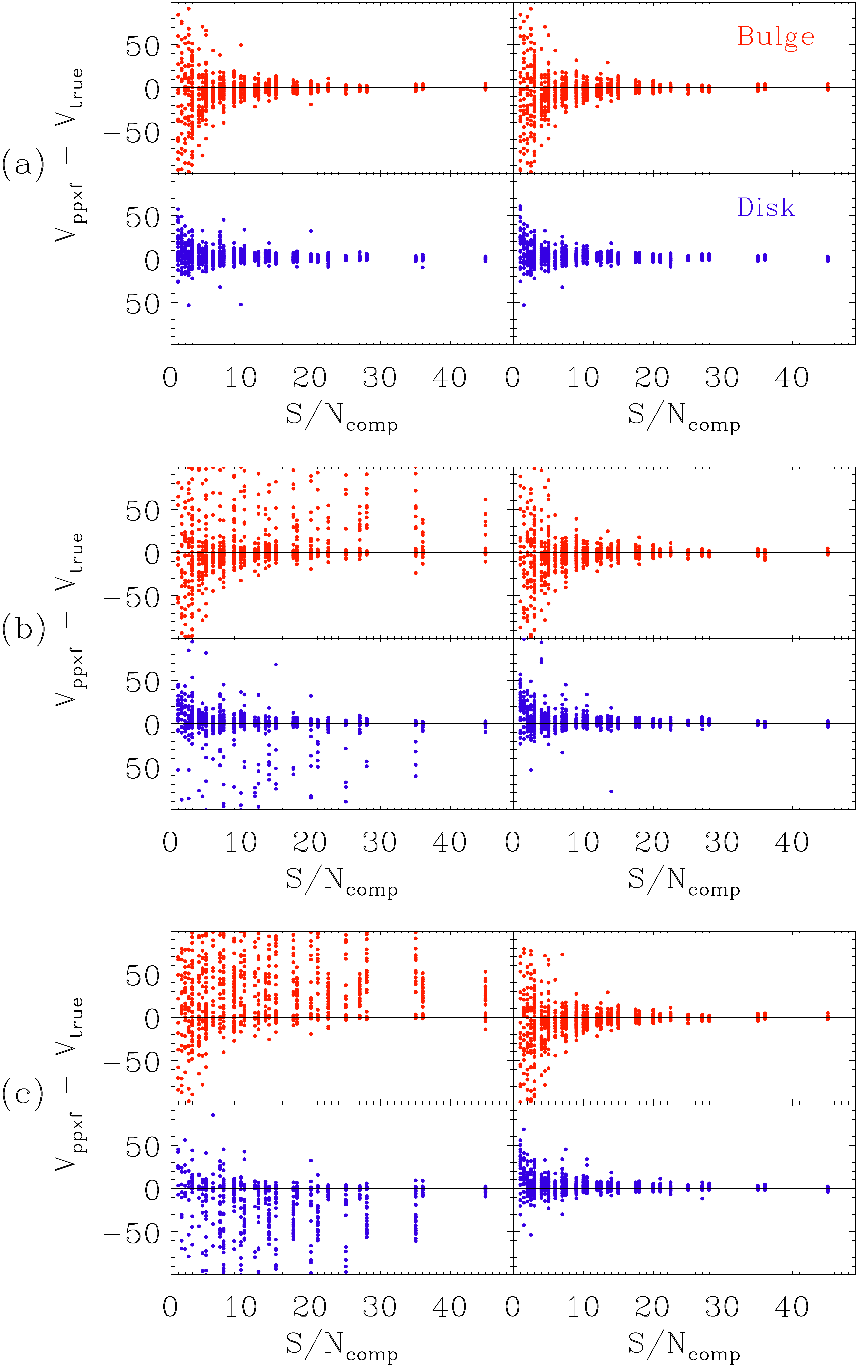}

\caption{The difference between the true velocities ($V_{\rm true}$) and the estimated solution from the pPXF ($V_{\rm ppxf}$) using three different initial guesses: (a)~initial guesses randomly chosen within $\pm$30\,km\,s$^{-1}$ of the true kinematics; (b)~initial guesses randomly chosen within $-200<V<200$\,km\,s$^{-1}$ and $0<\sigma<250$\,km\,s$^{-1}$; and (c)~the swapped initial guesses, where the true kinematics of the disk components have been used for the initial guesses of the bulge components, and vice versa. For good initial guesses, pPXF finds the true solution regardless of the use of the swapping routine. However, without the swapping routine, pPXF sometimes fails to find the true solution  when the initial guesses are significantly different from the true solution.}
\label{toy}
\end{figure}

\subsection{Testing the method}
We have generated 980 mock spectra with various input kinematics, B/T, and S/N in order to test the method described in the previous section. We adopted 12~Gyr and 1~Gyr model spectra from the MILES single stellar population models (Vazdekis et~al.\ 2010) for the bulge and disk components respectively. Two model spectra are shifted and convolved to yield the specified input kinematics (Figure~\ref{model}), and then the two components are weighted and summed according to the specified B/T ratio (in the range 0.1 to 0.9). Noise is then added to explore the dependence on S/N in the range 10 to 50~\AA$^{-1}$.

We then ran pPXF on the mock spectrum to test the swapping routine for three initial guesses: initial guesses randomly chosen within $\pm$30\,km\,s$^{-1}$ of the input kinematics; initial guesses randomly chosen in the ranges $-200<V<200$\,km\,s$^{-1}$ and $0<\sigma<250$\,km\,s$^{-1}$ for both components; and swapped initial guesses, where the true parameters of the disk kinematics are used as the initial guesses for the bulge kinematics, and vice versa. 

Figure~\ref{toy} shows the comparison between the true velocities ($V_{\rm true}$) and the outputs from the pPXF ($V_{\rm ppxf}$) with and without the swapping routine. When the input guesses are close enough to the true solution (Figure~\ref{toy}(a)), the swapping routine is mostly not activated, and both algorithms generate similar results. The result also shows that the error of the pPXF solution is highly dependent on the S/N of each component (S/N$_{\rm comp}$), calculated by multiplying the S/N by the weight of each component. Note that the error in the kinematics of each component correlates more tightly with S/N$_{\rm comp}$ than the overall S/N of the spectrum. When using the random initial guesses (Figure~\ref{toy}(b)), pPXF starts to show a significant error for some cases when the swapping routine is not used, whereas with the swapping routine it generates similar results as in case~(a). When the wrong initial guesses are used (Figure~\ref{toy}(c)), pPXF fails to estimate the true solution without the swapping routine even for the high S/N spectra. Although we only present the results for the rotation velocity here, we found the same results for the velocity dispersion. For observed spectra, there is no guarantee that the input guess will be close enough to the true solution to give a good outcome. In decomposing the bulge and disk kinematics it is therefore essential to break the dependence of the solution on the input guesses by using the swapping routine.

We have tested the bias in the kinematics due to the swapping routine and its constraint on $\sigma$. We chose one template from the MILES stellar population synthesis models. This single spectrum is shifted and convolved to have a rotation velocity of 100\,km\,s$^{-1}$ and a velocity dispersion of 100\,km\,s$^{-1}$. We then generated 1350 spectra by adding random noise to make the S/N in the range 10 to 30~\AA$^{-1}$. We tried to decompose these single-population and single-kinematics spectra assuming they are composed of two components with B/T ratio in the range 0.1 to 0.9. Both components are estimated to have similar kinematics with small bias. The median rotation velocities of bulge and disk components are, respectively, 98 and 101\,km\,s$^{-1}$ with a standard deviation of 8\,km\,s$^{-1}$. The median velocity dispersions of bulge and disk components are, respectively, 102 and 98\,km\,s$^{-1}$ with a standard deviation of 3\,km\,s$^{-1}$. Therefore, we conclude that the swapping routine does not artificially generate kinematical distinctness (i.e.\ pressure-supported bulges and rotation-supported disks). 

\begin{figure}
\centering
\includegraphics[width=0.47\textwidth]{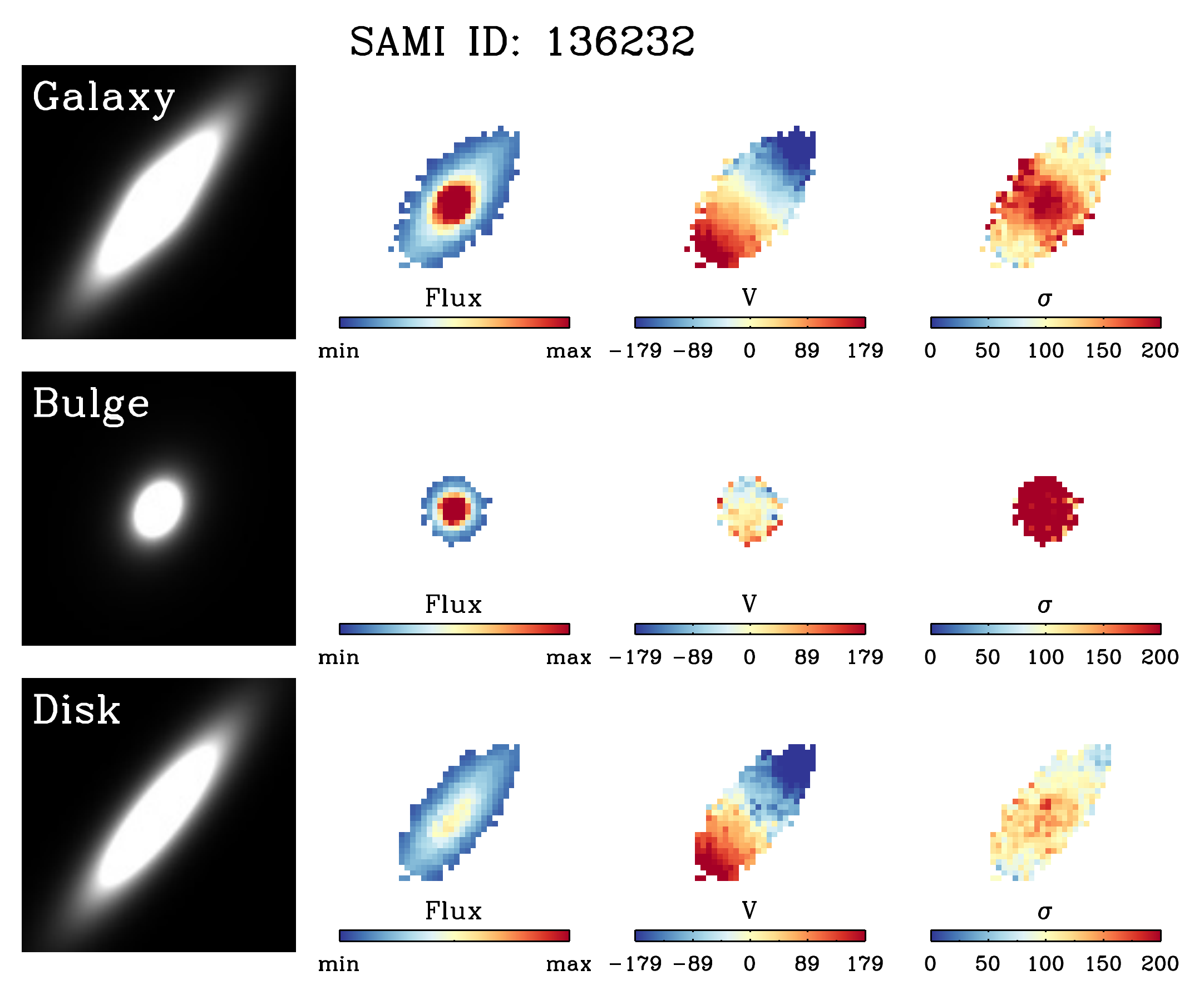}
\caption{Example two-dimensional flux and kinematic maps. The first column shows the flux model from the photometric decomposition. The other three columns present, respectively, the flux, rotation velocity and velocity dispersions provided by the SAMI kinematic maps. Note that for each component we only show SAMI spaxels whose S/N is greater than 3~\AA$^{-1}$. More examples can be found in Appendix~\ref{app:kin}.}
\label{kin}
\end{figure}

\section{Application}
\label{sec:app}
\subsection{Application to SAMI data}
We used SAMI internal data release v0.11 data cubes (Scott et~al.\ 2018). We have combined spectra from both the blue and red arms of SAMI. The red spectra have been broadened to have the same instrumental resolution as the blue spectra (2.65\,\AA) before combining them. The combined spectra have then been de-redshifted by applying $(1+z)^{-1}$ and rebinned on to a logarithmic wavelength scale with 57.9\,km\,s$^{-1}$ spacing using the {\sc{LOG$\_$REBIN}} routine provided by the pPXF package. 

The MILES stellar library (S\'{a}nchez-Bl\'{a}zquez et~al.\ 2006), consisting of 985 stellar spectra, has been used to provide templates for the full-spectral fitting. The template spectra, with a resolution of 2.5\,\AA, have been convolved to the same resolution as the SAMI spectra (2.65\,\AA). The templates are also normalised to maintain a mean flux of approximately unity, which is necessary for applying Equations~\ref{eq:fb1} and \ref{eq:fb2}. To extract the optimal template for each spaxel, we first generated three to five binned spectra from elliptical annuli following the light distribution of the galaxy constructed so that the S/N of each binned spectrum is at least 25\,\AA$^{-1}$. Annular binned spectra easily achieve our S/N requirement (25\,\AA$^{-1}$), while also accounting for strong radial gradients in stellar populations. The shape of annular bins is more regularised according to the radial light profile than that of Voronoi bins (Cappellari \& Copin 2003), and therefore can generate more representative templates to describe radial gradients in stellar population. See Scott et~al.\ (2018) and van de Sande et~al.\ (2017b) for more details on the spaxel binning scheme of SAMI. Then, we extracted the best-fit model spectra for each annulus from pPXF using the stellar templates.  

Using just the best-fit model spectra for these high-S/N annular spectra reduces the mismatch in templates for individual spaxels with low S/N. We have tested different combinations of the annular templates for bulge and disk components and found the lowest mean reduced $\chi^2$ when we use all the annular model spectra for the disk components and all the annular model spectra plus an additional template from the central spectra measured with a 2\arcsec-diameter aperture for the bulge components.

We used a 12th-order additive Legendre polynomial to reduce the impact of template mismatches on the measurement of kinematics (see Appendix~A.4 in van de Sande et~al.\ 2017b for more details on the choice of the order of this polynomial). We fit a Gaussian line-of-sight velocity distribution (LOSVD) to extract the rotation velocity and velocity dispersion. We applied an additional boundary condition of $\sigma_{\rm disk} < \sigma_{\rm galaxy}+ \sigma_{\rm error, galaxy}$, where $\sigma_{\rm galaxy}$ and $\sigma_{\rm error, galaxy}$ have been obtained from a pPXF run with a single component, which reduces the computing time and failure to converge. The pPXF run with a single component followed the steps described in van de Sande et~al.\ (2017b). For fitting  two components we set the {\sc fraction} keyword to be active and used the bulge fraction on each spaxel from the photometric bulge-disk decomposition. 

For each spaxel spectrum, we ran pPXF three times to obtain a better estimate of the input noise and determine which spectral pixels to include. First, we ran pPXF with uniform noise and `good' pixels defined by initial masking of emission lines and the gap between the blue and red spectra (Figure~\ref{ppxf}).
We then estimate the noise as the scaling of the original noise from SAMI using a ratio between the mean rms of the original noise and the mean rms of the residuals between the spectrum and the best fit. The second run was performed with the updated noise while activating the CLEAN keyword, which removes outliers using 3$\sigma$-clipping and updates the `good' pixels. The improved noise estimate from the first and second runs and the updated good pixels from the second run were then used for the final pPXF run.

\begin{figure}
\centering
\includegraphics[width=\columnwidth]{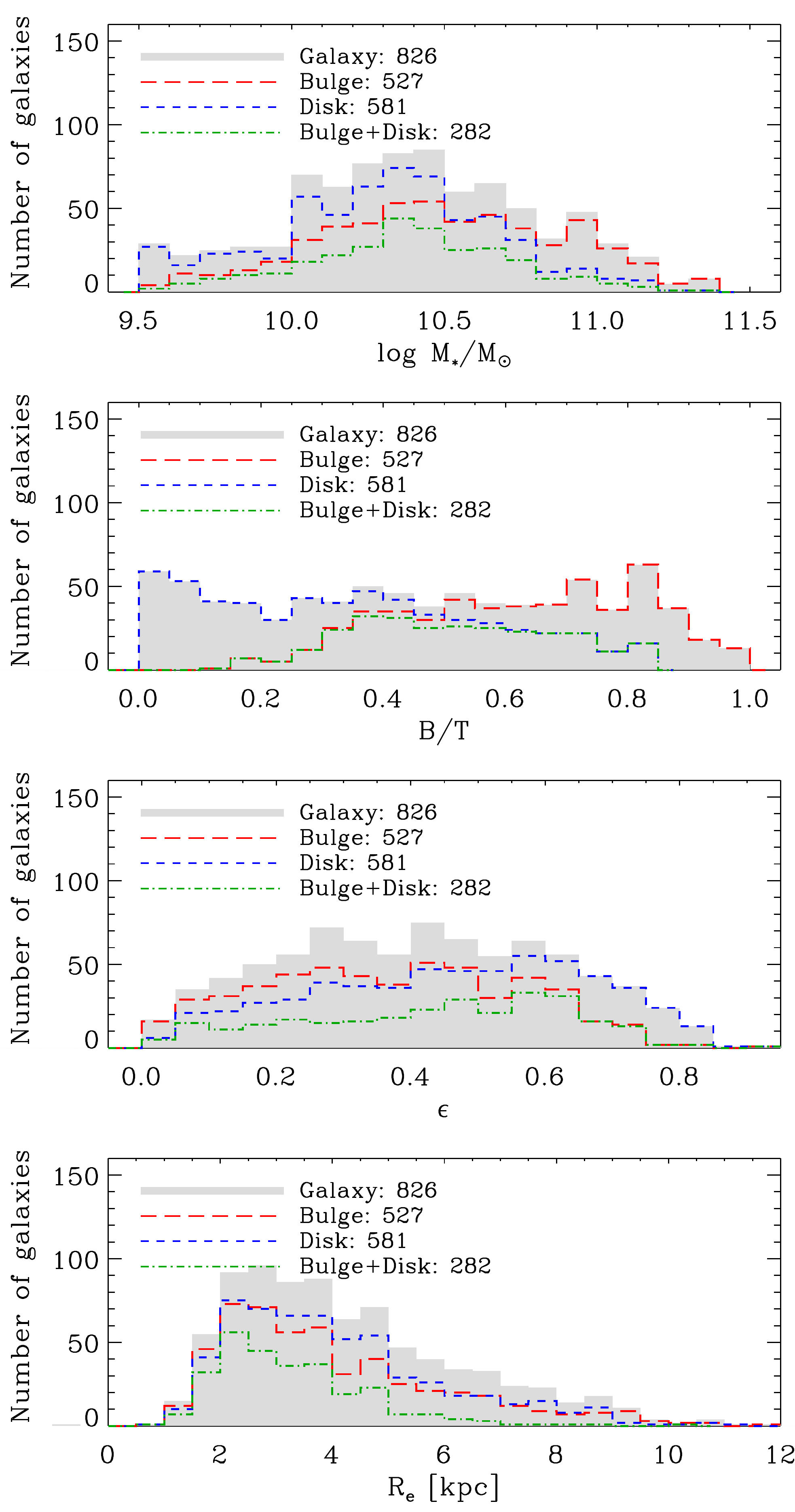}
\caption{The distribution of the total stellar mass ($M_*$), bulge-to-total ratio (B/T), ellipticity ($\epsilon$), and effective radius ($R_{\rm e}$) for the 826 galaxies in our sample. The distribution in B/T is nearly uniform, suggesting our sample includes similar numbers of a wide range of galaxy types.}
\label{sample}
\end{figure}

\begin{figure*}
\centering
\includegraphics[width=\textwidth]{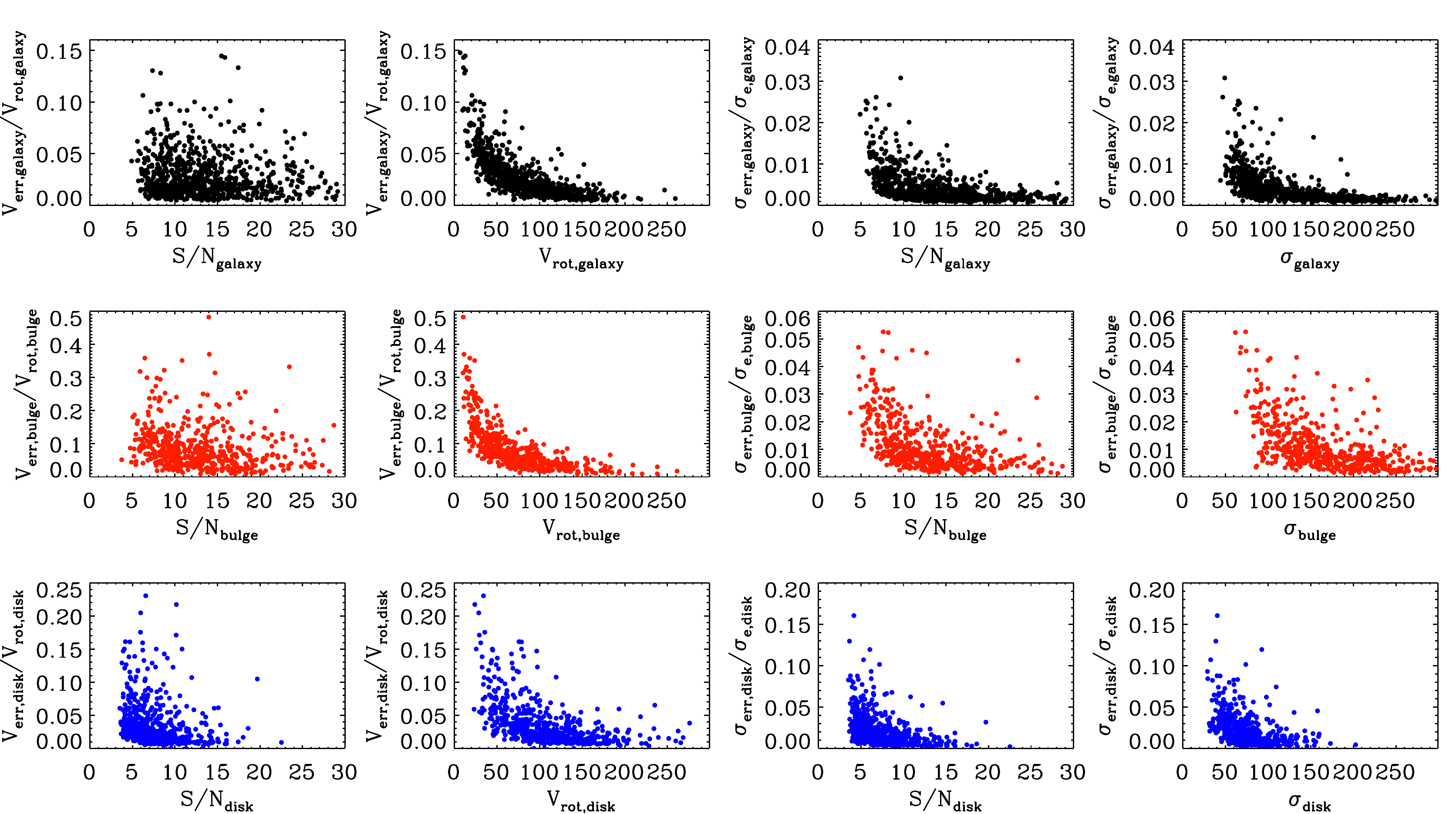}
\caption{The relative errors in stellar kinematics for the whole galaxy (top), the bulge component (middle), and the disk component (bottom). The relative errors increase at lower S/N and lower values of $V$ and $\sigma$.}
\label{err}
\end{figure*}   

Figure~\ref{kin} shows an example of the two-dimensional kinematic maps for the whole galaxy and for the bulge and disk components. We only show spaxels whose continuum S/N is greater than 3\,\AA$^{-1}$, with the S/N of bulge and disk components defined, respectively, as (S/N)$_{\rm bulge}$ = (B/T)$\times$(S/N) and (S/N)$_{\rm disk}$ = (1$-$B/T)$\times$(S/N). Note that S/N is measured over the wavelength range 4500--7000\,\AA. The bulge component map shows relatively little rotation but high velocity dispersions. The disk component map shows rotation velocities similar to the galaxy as a whole but lower velocity dispersions. More kinematics maps for various galaxy types can be found in Appendix~\ref{app:kin}. 

\subsection{Samples}
\label{sec:sam}
Below the spectrograph instrumental resolution, the precision and accuracy of stellar velocity dispersion measurements at fixed S/N decreases rapidly with decreasing velocity dispersion. The instrumental resolution of SAMI is $\sim$70\,km\,s$^{-1}$ for the blue spectra, from which the stellar velocity dispersions are measured. This corresponds to $\sim$10$^{9.5}$\,M$_{\odot}$ according to the stellar mass--velocity dispersion relation (Barat et~al.\ 2019). We therefore selected the 2169 galaxies with total stellar mass greater than $10^{9.5}\,M_{\odot}$ from the full sample of 2964 SAMI galaxies. The output from the photometric bulge-disk decomposition is available for 1400 galaxies (see Section~\ref{sec:phot}); 288 cluster galaxies in the southern hemisphere are missing due to the lack of SDSS imaging, and an additional 481 galaxies are excluded because either their bulge component is unresolved or their fit is ill-constrained. We do not find a bias in morphology (or B/T) for those 769 galaxies excluded from the sample. To reduce the effect of seeing, we excluded 66 galaxies whose galaxy $R_{\rm e}$ is smaller than 1.5 times the PSF half width half maximum (HWHM), which for the SAMI sample ranges is 1--1.5\arcsec. Additionally, we excluded 508 galaxies because more than 30\% of spaxels within the galaxy $R_{\rm e}$ have S/N$_{\rm bulge}$ and S/N$_{\rm disk}$ less than 3\,\AA$^{-1}$. This results in a final sample of 826 galaxies, of which 527 and 581 are available for measuring bulge and disk kinematics (respectively) because more than 70\% of spaxels within the galaxy $R_{\rm e}$ have S/N$_{\rm bulge}$ and S/N$_{\rm disk}$ greater than 3\,\AA$^{-1}$; 282 galaxies have kinematics available for both components. Note that the kinematics of the bulge and disk components have been measured using spectroscopic decomposition even if one component is ultimately excluded from the sample due to low S/N. Raising our S/N threshold to 7\,\AA$^{-1}$ instead of 3\,\AA$^{-1}$, which reduces the sample size by 55\%, has minimal impact on our results. 

In Figure~\ref{sample}, we present the distribution of galaxy mass, B/T, $\epsilon$, and $R_{\rm e}$ of our sample. The B/T of the sample is uniformly distributed, so we expect relatively little sample bias with galaxy morphology. Specifically, our sample includes 195 ellipticals, 336 lenticulars, and 295 spirals, of which 93\%, 79\%, and 27\% are available for bulge kinematics. On the other hand, 38\%, 70\%, and 92\% of respectively ellipticals, lenticulars, and spirals are available for measuring disk kinematics. The bulge components in this study are expected to be `classical' bulges, and most of the `pseudo' bulges are excluded by the sampling criteria. Only 23 bulges in this study are from late-type galaxies with bulge S\'ersic index $n < 1.5$, as obtained from the photometric bulge-disk decompositions (Section~\ref{sec:phot}). Note that our sample still includes 116 late-type galaxies which are suspected to host `pseudo' bulges (bulge S\'ersic $n < 1.5$), but only 20\% of them are available to measure bulge kinematics due to the S/N limit.

\section{Results}
\label{sec:res}
\subsection{Measurement of stellar kinematics}
\label{sec:mkin}
We measure the stellar kinematics of galaxies within 1\,$R_{\rm e}$ using the pPXF fitting with a single component. The stellar kinematics of the bulge and disk components are also measured within each galaxy's $R_{\rm e}$ for direct comparison, but using the pPXF fitting for two components. Spaxels have been excluded from the measurement of a given component when the S/N of that component is lower than 3\,\AA$^{-1}$ or when V and $\sigma$ of that component are greater than 400 and 600\,km\,s$^{-1}$, respectively. We present a comparison of the mean $\chi^2$ distribution between single- and two-component pPXF fitting in Appendix~\ref{app:chi}. The rotation velocity ($V_{\rm rot}$) has been estimated using the velocity width ($W \equiv V_{90} - V_{10}$; Catinella et~al.\ 2005), which is the difference between the 90th and 10th percentile points of the histogram of the rotation velocities within 1\,$R_{\rm e}$ (see also Cortese et~al.\ 2014 and Barat et~al.\ 2019). Then $V_{\rm rot}$ has been calculated as 
\begin{equation}
\label{eq:vel}
V_{\rm rot} \equiv \frac{W}{2(1+z)\sin(i)} ~,
\end{equation}
where $i$ is the inclination derived using the axial ratio (b/a):
\begin{equation}
\cos(i) = \sqrt{\frac{(b/a)^2-q_0^2}{1-q_0^2}} ~.
\end{equation}
We assumed the intrinsic flatness ($q_0$) to be 0.2 for galaxies with B/T~$<0.4$ and 0.6 for the others. We measured the velocity dispersion ($\sigma_{\rm e}$) as the flux-weighted mean of velocity dispersions of spaxels within 1\,$R_{\rm e}$,
\begin{equation}
\label{eq:sig}
\sigma_{\rm e}^2 \equiv \frac{\sum_i F_i\sigma_i^2}{\sum_i F_i} ~,
\end{equation}
where $\sigma_i$ and $F_i$ are the velocity dispersion and the continuum flux in each spaxel. Note that $F_i$ for bulge and disk components are relative fluxes derived using B/T. Our definition of $\sigma_{\rm e}$ differs from the one measured using aperture spectra which is an approximation to the second velocity moment. We exclude the impact of rotation velocity from our $\sigma_{\rm e}$ to consider only dispersion components. Aperture correction has not been applied to our measurements because the correction reduces the velocity dispersion by at most 1\%.

Errors in the rotation velocity and velocity dispersion ($V_{\rm err}$ and $\sigma_{\rm err}$) have been estimated with a Monte Carlo method: we randomly generated $V_{\rm rot}$ and $\sigma$ for each spaxel based on the value and error from pPXF and obtained the standard deviation in the derived measurements of $V_{\rm rot}$ and $\sigma_{\rm e}$ from 100 repeats. The errors are typically a few km\,s$^{-1}$ and correlate with the S/N of each component (Figure~\ref{err}). We present a comparison between the error estimates from the Monte Carlo and noise-shuffling methods in Appendix~\ref{app:err}. 

\begin{figure*}
\centering
\includegraphics[width=\textwidth]{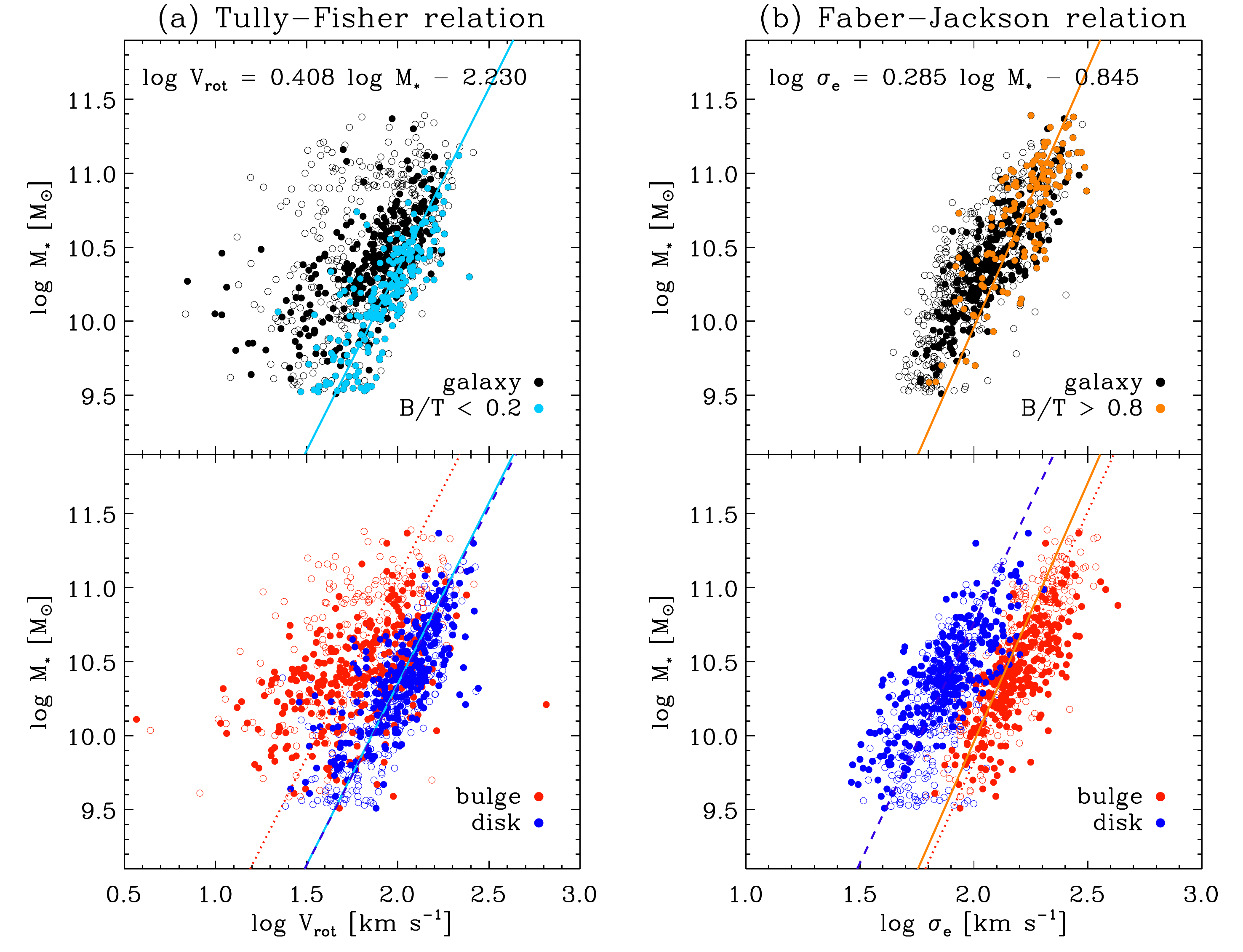}
\caption{The scaling relations with stellar mass: (a)~the Tully-Fisher relation ($\log M_{*}$--$\log V_{\rm rot}$) and (b)~the Faber-Jackson relation ($\log M_{*}$--$\log \sigma_{\rm e}$) for the whole galaxy (top panels) and the bulge and disk components (bottom panels). Note that $M_{*}$ is total stellar mass, not the mass of either component. Open circles indicate galaxies having a kinematic measurement for just one component and filled circles show the 282 galaxies with measurements for both components. The fiducial relations for the Tully-Fisher relation (cyan line) and the Faber-Jackson relation (orange line) have been derived, respectively, from the single-component fits for galaxies with B/T$ < 0.2$ (cyan points) and B/T$ > 0.8$ (orange points). The dashed and dotted lines show (respectively) the fit for bulge and disk components.}
\label{scaling}
\end{figure*}
     
\begin{figure*}
\centering
\includegraphics[width=\textwidth]{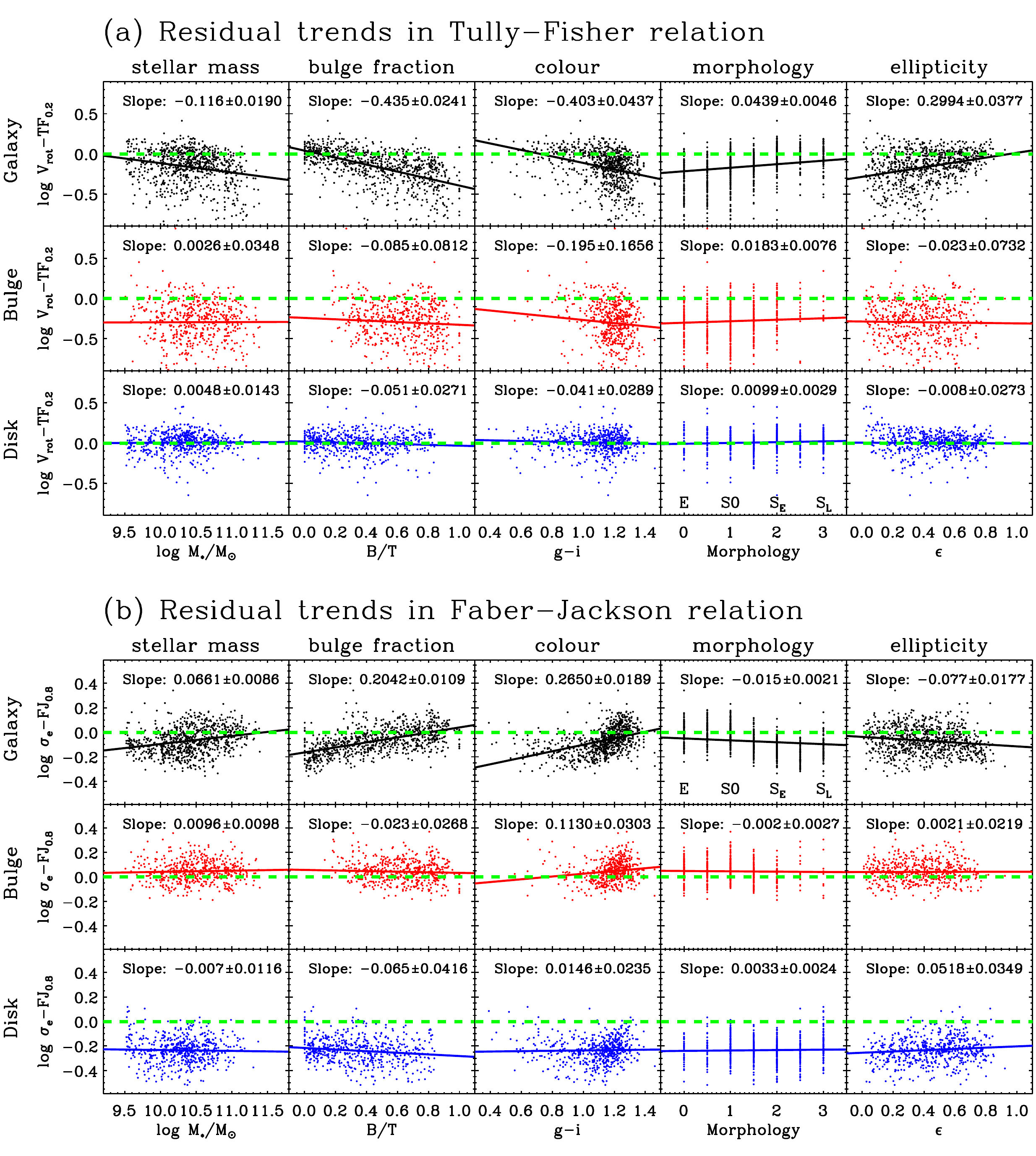}
\caption{The residual trends in (a)~the Tully-Fisher relation and (b)~the Faber-Jackson relation as functions of galaxy stellar mass, bulge fraction, colour, morphology, and ellipticity. The solid lines show the least-squares fit to the trend of the residuals with each galaxy property; the dashed lines show the fiducial relations (cyan and orange lines in Figure~\ref{scaling}). The slope from the least-squares fit is given in each panel. For single-component galaxy kinematic measurements (top rows), both relations show highly significant residual correlations with all galaxy properties, whereas, for two-component bulge and disk measurements (middle and bottom rows), neither relation shows a highly significant residual correlation with any galaxy property (only the bulge Faber-Jackson relation with colour shows a marginally significant trend).}
\label{residual}
\end{figure*}    

 \begin{table*}
\centering
\caption{Fits to the residuals from the Tully-Fisher and Faber-Jackson relations}
\begin{tabular}{lllrrr}
\hline\hline
  & && Galaxy &  Bulge & Disk \\
\hline
\multirow{6}{*}{Tully-Fisher relation}  & \multirow{3}{*}{$V_{\rm rot}$ offset$^a$} &All sample& $-0.116\pm0.010$ & $-0.260\pm0.014$ & $+0.016\pm0.007$ \\
& &{Early-type} & $-0.186\pm0.013$ & $-0.280\pm0.015$ & $+0.004\pm0.009$ \\
& &{Late-type}  & $+0.007\pm0.011$ & $-0.154\pm0.031$ & $+0.034\pm0.009$ \\
 & \multirow{3}{*}{$V_{\rm rot}$ scatter$^b$} & All sample & $0.125\pm0.007$ & $0.171\pm0.009$ & $0.070\pm0.005$ \\
& & Early-type& $0.121\pm0.009$ & $0.166\pm0.009$ & $0.069\pm0.006$ \\
& & Late-type & $0.072\pm0.008$ & $0.137\pm0.019$ & $0.061\pm0.006$ \\
\hline
 \multirow{6}{*}{Faber-Jackson relation} & \multirow{3}{*}{$\sigma_{\rm e}$ offset$^c$}& All sample & $-0.073\pm0.005$ & $+0.039\pm0.005$ & $-0.238\pm0.005$ \\
 & &{Early-type}& $-0.034\pm0.005$ & $+0.046\pm0.005$ & $-0.227\pm0.007$ \\
 & &{Late-type}& $-0.149\pm0.006$ & $-0.006\pm0.014$ & $-0.244\pm0.008$ \\
&  \multirow{3}{*}{$\sigma_{\rm e}$ scatter$^d$} & All sample & $0.074\pm0.003$ & $0.058\pm0.003$ & $0.061\pm0.003$ \\
&& Early-type & $0.066\pm0.003$ & $0.055\pm0.003$ & $0.063\pm0.005$ \\
&& Late-type & $0.057\pm0.004$ & $0.053\pm0.009$ & $0.057\pm0.005$ \\
\hline\hline
\multicolumn{5}{l}{$^a$ The median of ($\log V_{\rm rot} - $TF$_{0.2}$)}\\
\multicolumn{5}{l}{$^b$ The median absolute deviation from ($\log V_{\rm rot} - $TF$_{0.2}-V_{\rm rot}$ offset)}\\
\multicolumn{5}{l}{$^c$ The median of ($\log \sigma_{\rm e} - $FJ$_{0.8}$)}\\
\multicolumn{5}{l}{$^d$ The median absolute deviation from ($\log \sigma_{\rm e} - $FJ$_{0.8}- \sigma_{\rm e}$ offset)}\\
\end{tabular}
\label{tab:scatter} 
\end{table*}
 
\subsection{Tully-Fisher ($M_*$--$V_{\rm rot}$) relation}
The Tully-Fisher relation, a relatively tight correlation between stellar mass (or luminosity) and rotation velocity, has been regarded as largely applying to late-type galaxies. In Figure~\ref{scaling}(a), we define the fiducial Tully-Fisher relation (TF$_{0.2}$) using single-component kinematic fits to late-type galaxies with B/T~$<0.2$ which is fairly consistent with the usual Tully-Fisher relation ($M_{*} \propto V_{\rm rot}^{2.5}$):
\begin{equation}
\log V_{\rm rot} = 0.408 \log M_{\rm *} - 2.230.
\end{equation}
We indeed find early-type galaxies with higher B/T have single-component $V_{\rm rot}$ that are substantially scattered toward lower values than the fiducial relation would predict (see also Cortese et~al.\ 2014; Oh et~al.\ 2016; Aquino-Ort$\acute{\rm i}$z et~al.\ 2018; Barat et~al.\ 2019). Based on the decomposed kinematics, we found a tight correlation between total $M_{*}$ and $V_{\rm rot}$ for disk components that is close to the fiducial relation based on low B/T galaxies (though with a slight offset to higher $V_{\rm rot}$ at given $M_{*}$). By contrast, bulge components have much more scatter to low $V_{\rm rot}$ at given $M_{*}$ than the fiducial relation. 

We investigated the residual in $V_{\rm rot}$ in Figure~\ref{residual}(a) and Table~\ref{tab:scatter}. The fiducial relation TF$_{0.2}$ has been subtracted from the measured $\log V_{\rm rot}$ for the same stellar mass. The magnitude of the residual in $V_{\rm rot}$ becomes larger as the galaxy properties tend toward those of early types: galaxies that are more massive, bulge-dominated, redder, and having an earlier morphology deviate more from the fiducial relation. For example, early-type galaxies show a mean $V_{\rm rot}$ offset of $-0.186$\,dex from TF$_{0.2}$ (Table~\ref{tab:scatter}). On the other hand, the disk components do not show a trend in the $V_{\rm rot}$ residual with respect to any of these galaxy properties; even the disk components from early-type galaxies (e.g.\ B/T~$> 0.4$) show nearly zero residuals (see also Table~\ref{tab:scatter}), implying that the disk components of early-type galaxies share the same Tully-Fisher relation as low B/T late-type galaxies. The $V_{\rm rot}$ of the bulge components shows a significant offset from the fiducial relation and a substantially larger scatter in $V_{\rm rot}$. The bulge components from late-type galaxies tend to show smaller $V_{\rm rot}$ offsets from TF$_{0.2}$ (mean offset of $-0.154$\,dex) than those from early type galaxies (mean offset of $-0.280$\,dex), implying that the bulges of late-type galaxies are more rotationally supported than those of early-type galaxies.

The Tully-Fisher scaling relation between mass and rotation velocity appears to apply to the rotating (i.e.\ disk) component of all types of galaxies. An increasing bulge fraction dilutes the relation, hiding it completely in early-type galaxies until they are decomposed.

\subsection{Faber-Jackson ($M_{*}$--$\sigma_{\rm e}$) relation}
A similar analysis has been applied to the Faber-Jackson relation between stellar mass (or luminosity) and velocity dispersion (Figure~\ref{scaling}(b)). The scatter in the Faber-Jackson relation is less prominent than that from the Tully-Fisher relation, even including late-type galaxies. The fiducial Faber-Jackson relation (FJ$_{0.8}$) has been defined using bulge-dominated galaxies with B/T~$> 0.8$ which is also comparable to the usual Faber-Jackson relation ($M_{*} \propto \sigma_{\rm e}^{3.5}$):
\begin{equation}
\log \sigma_{\rm e} = 0.285 \log M_{\rm *} - 0.845.
\end{equation}
 The Faber-Jackson relation of the bulge component is very similar to the fiducial relation, although with a slight offset to higher $\sigma_{\rm e}$ at fixed $M_{*}$. Remarkably, there is a well-defined Faber-Jackson relation, almost parallel to FJ$_{0.8}$, linking the velocity dispersion of disk components to their galaxies' stellar masses, although the disk velocity dispersion at fixed $M_{*}$ is about a factor of 2 lower than the fiducial relation would predict. The pressure-supported bulge components are expected to show a tight correlation between the stellar mass and velocity dispersion, but it is surprising to find a similarly tight Faber-Jackson relation for rotation-supported disk components.

Although we excluded small galaxies ($R_{\rm e} < 1.5$ PSF HWHM) from our sample, we might still expect some effect of beam-smearing on our measurements of $\sigma_{\rm e}$. However, in Appendix~\ref{app:beam} we investigate the impact of beam-smearing on the Faber-Jackson relation of the disk components using annular kinematic measurements (excluding the centre where the beam-smearing is maximised) and find no substantive difference to the results presented here. Also, we found consistent results when we exclude components whose effective radius (of either component) is smaller than 2 PSF HWHM, which reduces the number of bulge components by 40\% and disk components by 3\%.

As with the Tully-Fisher relation, any residual trend in the Faber-Jackson relation with galaxy properties has disappeared with the separation of bulge and disk kinematics. In Figure~\ref{residual}(b) and Table~\ref{tab:scatter}, the residual in $\sigma_{\rm e}$ has been calculated by subtracting the fiducial relation FJ$_{0.8}$ from the observed $\log\sigma_{\rm e}$ for the same stellar mass. The magnitude of the residual in $\sigma_{\rm e}$ becomes larger as the galaxy properties tend toward those of late types: galaxies that are less massive, disk-dominated, bluer, and having a later morphology deviate more from the fiducial relation. Bulge components show a slightly larger $\sigma_{\rm e}$ (by 0.04\,dex) than FJ$_{0.8}$, while disk components show smaller $\sigma_{\rm e}$ (by about 0.24~dex) than FJ$_{0.8}$ (Table~\ref{tab:scatter}). When we separate early- and late-type galaxies, we find that their disk components have statistically consistent offsets and scatters (last column of Table 1). For bulges, we find the scatters are consistent, but the offset for late-type bulges is marginally less than that of the full sample ($-0.006\pm0.014$ vs. $0.039\pm0.005$, three standard deviations). Neither component shows a clear dependence of the $\sigma_{\rm e}$ residuals on any galaxy properties (Figure~\ref{residual}(b)). Note that 77\% of our sample have $10\lesssim \log M_*/M_{\odot} \lesssim 11$, and therefore, our results mainly hold for galaxies of intermediate stellar mass.

\subsection{The spin parameter $\lambda_{\rm R}$ and $V/\sigma$}
\label{sec:lamb}
\begin{figure}
\centering
\includegraphics[width=\columnwidth]{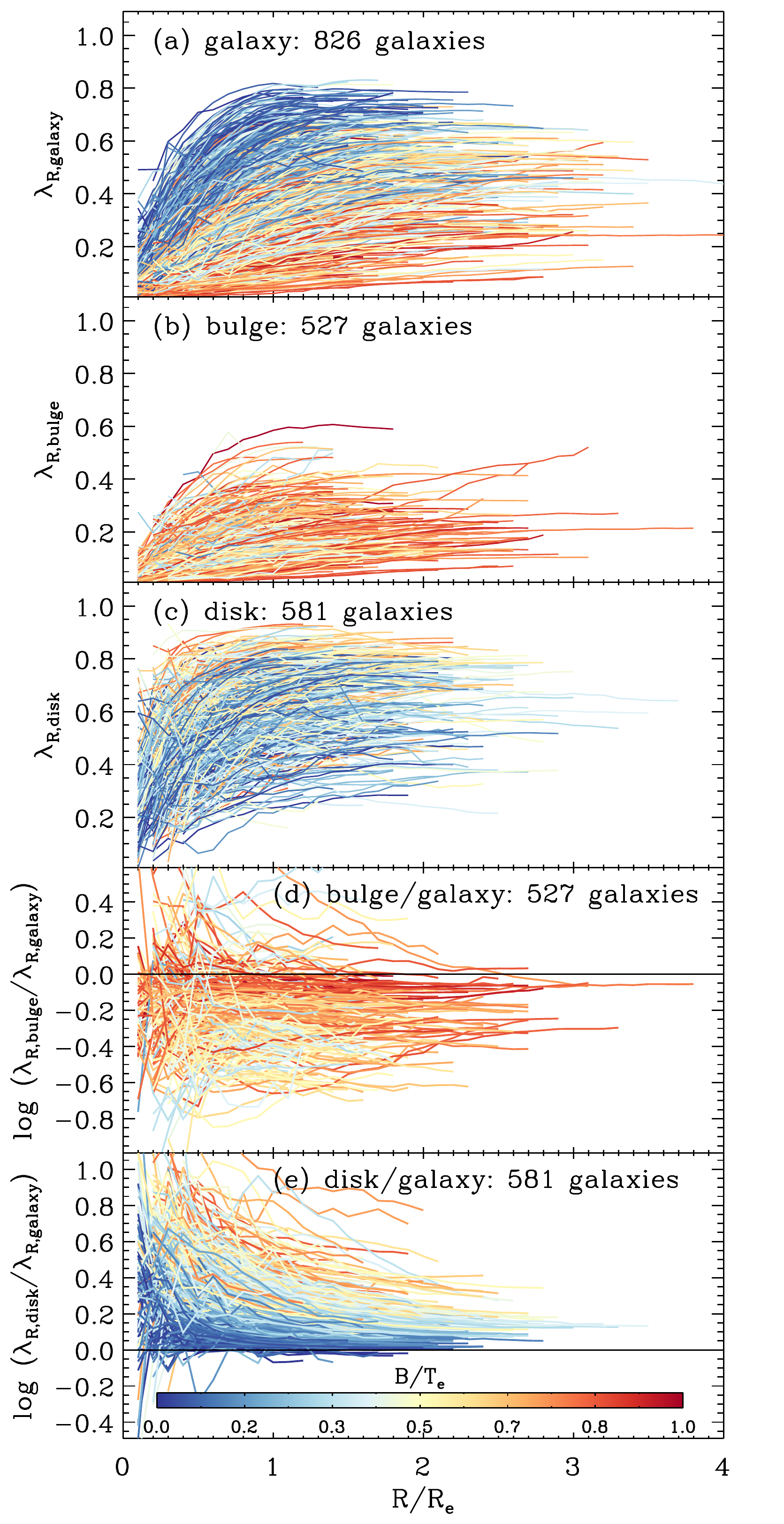}
\caption{$\lambda_{\rm R}$ profiles for (a)~galaxies, (b)~bulge components, and (c)~disk components, and the log of the ratio of the $\lambda_{\rm R}$ profile with respect to the whole galaxy of the (d)~bulge components and (e)~disk components. Galaxies show a clear dependence of $\lambda_{\rm R}$ profile on B/T$_{\rm e}$. The $\lambda_{\rm R}$ profiles of galaxies with low and high B/T$_{\rm e}$ show more discrepancy relative to the $\lambda_{\rm R}$ profiles of the bulge and disk components respectively.}
\label{lambp}
\end{figure}   

The spin parameter $\lambda_{\rm R}$ has been widely measured from integral field spectroscopy to quantify the dominance of the ordered or random motions in galaxies (e.g.\ Emsellem et~al.\ 2007, 2011; Cappellari 2016; van de Sande et~al.\ 2017b).  We have measured cumulative $\lambda_{\rm R}$  profiles using elliptical apertures (with position angle and ellipticity $\epsilon$ fixed at 1\,$R_{\rm e}$) following Emsellem et~al.\ (2007):
\begin{equation}
\lambda_{\rm R} \equiv \frac{\left\langle R\left| V \right| \right\rangle}{\left\langle R\sqrt{V^2+\sigma^2} \right\rangle} = \frac{\sum_i F_{i} R_{i}\left| V_{i} \right|}{\sum_i F_{i} R_{i}\sqrt{V^{2}_{i}+\sigma^{2}_{i}}} ~,
\end{equation}
 where $F_i$, $R_i$, $V_i$, and $\sigma_i$ are, respectively, the (relative) continuum flux, radius, mean velocity, and velocity dispersion of each spaxel. For direct comparison, the $\lambda_{\rm R}$ of the bulge and disk components are also measured using the same apertures as for whole galaxies. 

 We present the radial profile of $\lambda_{\rm R}$ for each component in Figure~\ref{lambp}. The $\lambda_{\rm R}$ increases as the aperture size increases, with most of the increase occurring within 1\,$R_{\rm e}$ for disks and disk-dominated galaxies and within 2\,$R_{\rm e}$ for bulges and bulge-dominated galaxies. At a given radius with a unit of galaxy's $R_{\rm e}$, the $\lambda_{\rm R}$ values of the bulge and disk components tend, respectively, toward the lower and higher ends of the distribution. The difference between the bulge and disk components in the $\lambda_{\rm R}$ is not only shown at a specific radius but at all radii, though the two distributions are rather broad and have some overlap. Galaxies show a clear dependence of $\lambda_{\rm R}$ on B/T$_{\rm e}$, whereas bulge and disk components do not show any clear relationship between $\lambda_{\rm R}$ and B/T$_{\rm e}$. In other words, the overall spin parameter of a galaxy is a direct result of the relative dominance of the bulge or disk component, while the spin parameter of the bulge and disk components is largely independent of which dominates.

The magnitude of the difference in $\lambda_{\rm R}$ depends on B/T$_{\rm e}$. Bulges from galaxies with high B/T$_{\rm e}$ show small differences with galaxies in $\lambda_{\rm R}$, whereas the discrepancy in $\lambda_{\rm R}$ is large for bulges with low B/T$_{\rm e}$ at a given radius (Figure~\ref{lambp}(d)). Similarly, the discrepancy in $\lambda_{\rm R}$ between galaxies and disks becomes more significant at higher B/T$_{\rm e}$ (Figure~\ref{lambp}(e)). Disk components exhibit larger $\lambda_{\rm R}$ relative to galaxies mainly because they have a lower velocity dispersion than galaxies, especially near the centre where galaxies are dominated by bulge components. 

\begin{figure}
\centering
\includegraphics[width=\columnwidth]{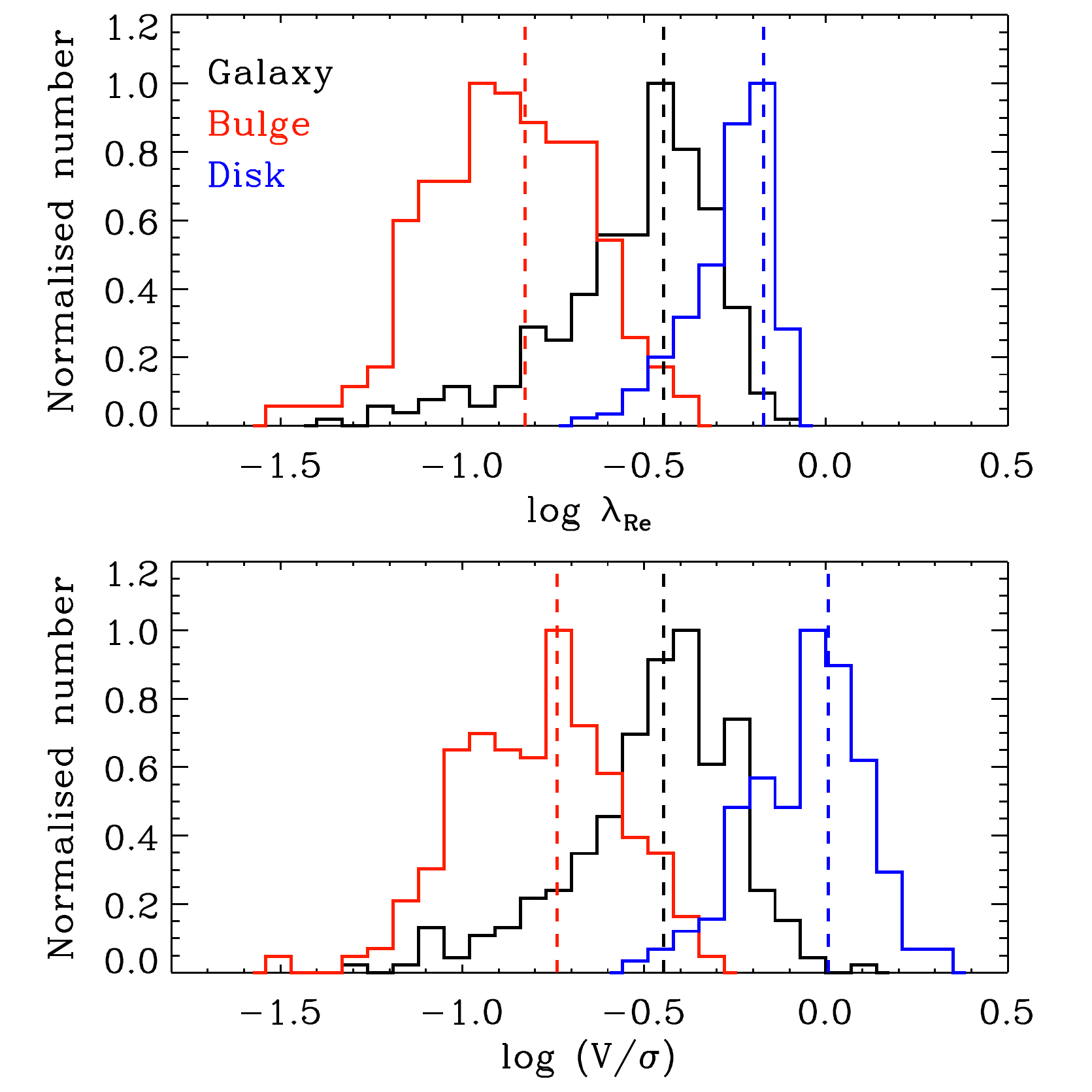}
\caption{The distribution of $\lambda_{\rm R_e}$ and $V/\sigma$ for 282 galaxies with both bulge and disk components. Black, red and blue histograms show, respectively, galaxies and their bulge and disk components; the dashed lines show the median of each distribution.}
\label{vsig}
\end{figure}    

As expected from the $\lambda_{\rm R}$ profile, the bulge and disk components show a clear separation in $\lambda_{\rm R_e}$ and $V/\sigma$ (Figure~\ref{vsig}). The $V/\sigma$ was measured as the flux-weighted mean within 1\,$R_{\rm e}$ following Cappellari et~al.\ (2007): 
\begin{equation}
\left(\frac{V}{\sigma}\right)^2 \equiv \frac{\langle V^2\rangle}{\langle\sigma^2\rangle} = \frac{\sum_i F_i V_{i}^2}{\sum_i F_i \sigma_i^2} ~.
\end{equation}
The figure only shows the 282 galaxies with enough S/N to measure the kinematics of both components satisfying our sampling criterion (Section~\ref{sec:sam}). Our sample galaxies show a broad range in $\lambda_{\rm R_e}$ and $V/\sigma$. On the other hand, the bulge and disk components show, respectively, low and high values in $\lambda_{\rm R_e}$ and $V/\sigma$; moreover, the $\lambda_{\rm R_e}$ and $V/\sigma$ distributions of the two components are well separated with little overlap. Note that the 282 galaxies in this analysis mostly have $10\lesssim \log M_*/M_{\odot} \lesssim 11$ and $0.3\lesssim$~B/T~$\lesssim0.7$ (Figure~\ref{sample}). The $\lambda_{\rm R_e}$ and $V/\sigma$ distributions of galaxies become broader, but both bulge and disk components show similar distributions when including all 826 galaxies.

\begin{figure}
\centering
\includegraphics[width=\columnwidth]{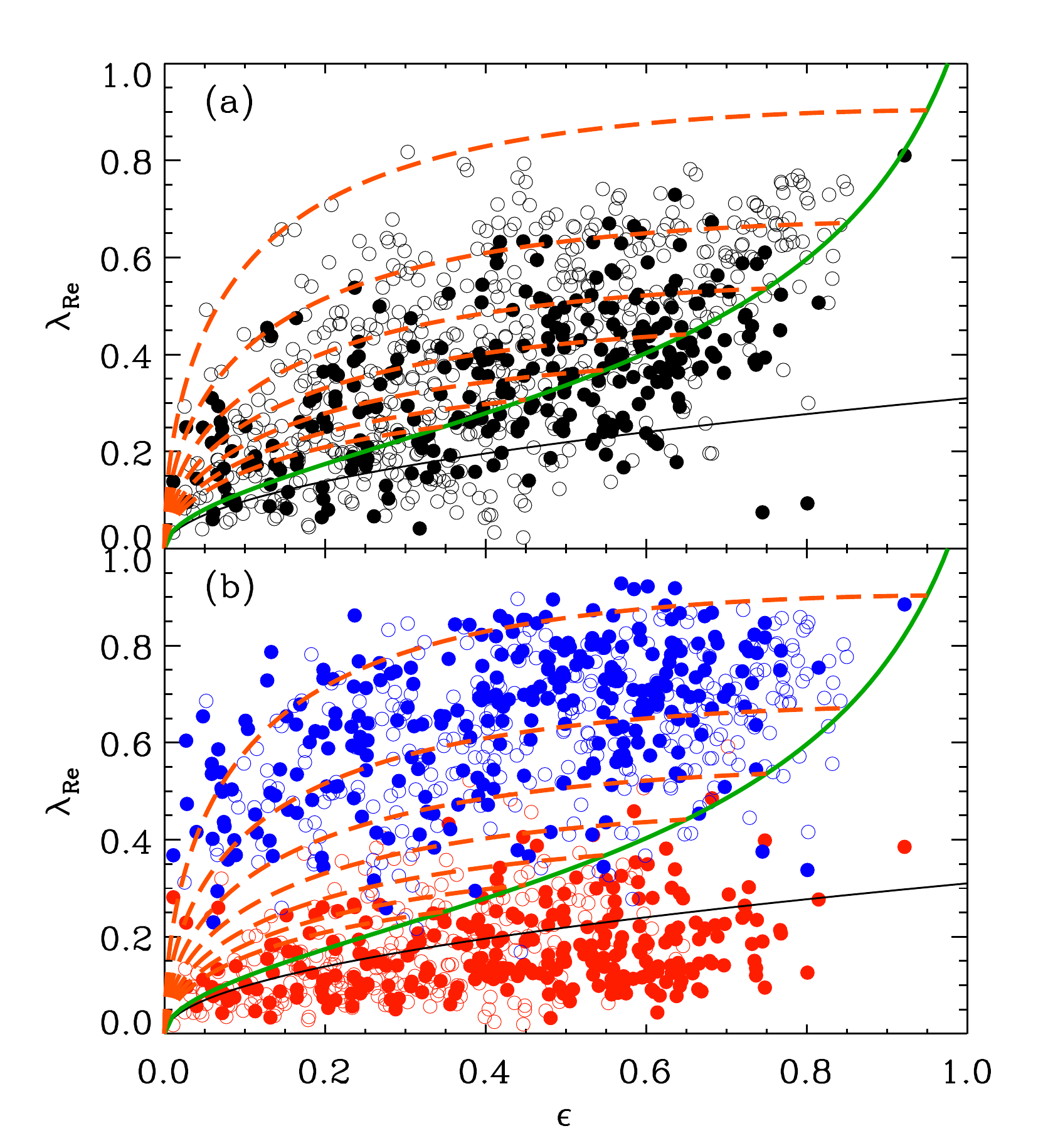}
\caption{The $\lambda_{\rm R_e}$--$\epsilon$ distribution for (a)~galaxies (black), (b)~the bulge (red) and disk (blue) components. The spin parameter $\lambda_{\rm R_e}$ for both components is also measured at 1\,$R_{\rm e}$ for the galaxy as a whole; likewise, $\epsilon$ from the galaxy is used for both components. Open circles are galaxies with a kinematic measurement for just one component, and filled circles are the 282 galaxies with measurements for both components. The black curve corresponds to $\lambda_{\rm R_e} = 0.31 \sqrt{\epsilon}$, which is often used to separate fast and slow rotators (Emsellem et~al.\ 2011). The green curve shows the theoretical expectation for the edge-on view of axisymmetric galaxies with $\beta = 0.7 \epsilon_{\rm int}$, where $\epsilon_{\rm int}$ is the intrinsic ellipticity (Cappellari et~al.\ 2007). The orange dashed curves correspond to the locii of galaxies with different intrinsic ellipticities $\epsilon_{\rm int}$ = 0.35--0.95 in steps of 0.1 (Emsellem et~al.\ 2011).}
\label{lamb}
\end{figure}  

\begin{figure}
\centering
\includegraphics[width=\columnwidth]{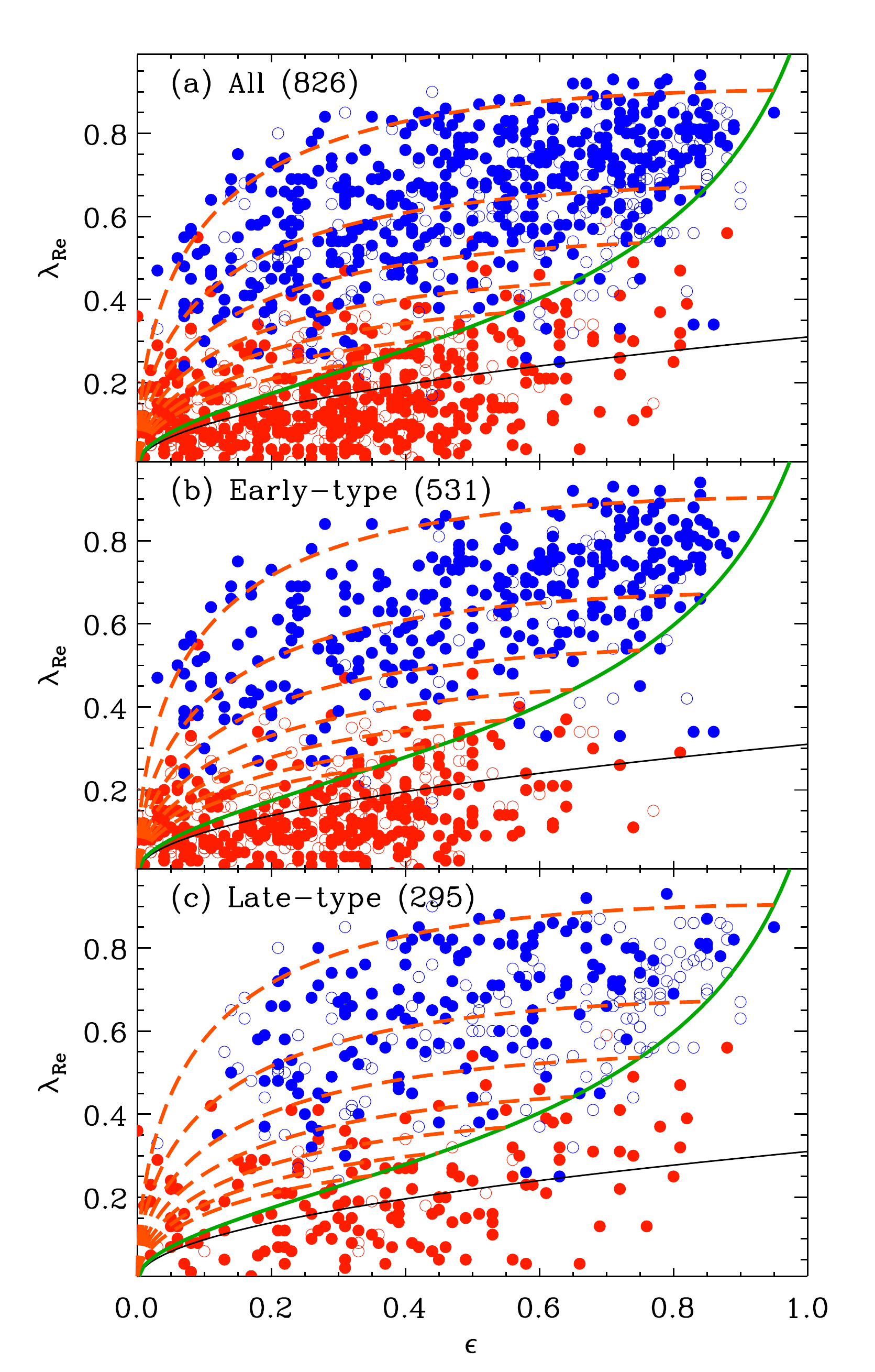}
\caption{The $\lambda_{\rm R_e}$--$\epsilon$ distribution for the bulge (red) and disk (blue) components from (a)~all morphological type, (b)~early-type, and (c)~late-type galaxies. The spin parameter $\lambda_{\rm R_e}$ for both components is measured at 1\,$R_{\rm e}$ for each component; likewise, $\epsilon$ from each component is used. Note that the use of $R_{\rm e}$ of each component allows us to have more galaxies available for measuring kinematics of both components. Out of the 826 sample galaxies, 664 and 633 are available for measuring bulge and disk components (respectively); 468 galaxies have kinematics available for both components (filled circles). The details of theoretical lines are the same as described in Figure~\ref{lamb}.}
\label{lambmor}
\end{figure}  
     
 The $\lambda_{\rm R_e}$--$\epsilon$ plane is often used as a diagnostic for fast and slow rotators (Emsellem et~al.\ 2011). Most early-type galaxies are fast rotators, implying they have significant rotating components (e.g.\ Emsellem et~al.\ 2007, 2011; Cappellari et~al.\ 2007; Krajnovi$\acute{\rm c}$ et~al.\ 2013; van de Sande et~al.\ 2017b). We present the $\lambda_{\rm R_e}$--$\epsilon$ distributions for galaxies and the bulge and disk components in Figure~\ref{lamb}. Note that $\lambda_{\rm R_e}$ for the bulge and disk components is measured at a radius of 1\,$R_{\rm e}$ for the galaxy as a whole to facilitate direct comparisons; likewise, $\epsilon$ from the galaxy is used for both components. Our sample, spanning a variety of galaxy types, is also dominated by fast rotators (87\%, 720/826) based on the demarcation line $\lambda_{\rm R_e} = 0.31 \sqrt{\epsilon}$ (Figure~\ref{lamb}(a); Emsellem et~al.\ 2011). On the other hand, 62\% (326/527) of the bulge components are classified as slow rotators, and the others are also found close to the demarcation line (Figure~\ref{lamb}(b)). We found only one galaxy whose disk component has $\lambda_{\rm R_e}$ below the demarcation line. 

In Figure~\ref{lambmor}, we present the $\lambda_{\rm R_e}$--$\epsilon$ distribution for early- and late-type galaxies. For the bulge (disk) components, $\lambda_{\rm R_e}$ is measured within the effective radius of the best-fit bulge (disk) model. Note that the use of $R_{\rm e}$ of each component gives more galaxies with kinematics for both components: 664 and 633 are available for measuring bulge and disk components (respectively), and 468 galaxies have kinematics available for both components. We find 58\% (384/664) of bulges from all morphological types are classified as slow rotators using $\lambda_{\rm R_e}$ measured at $R_{\rm e}$ of the bulge component (Figure~\ref{lambmor}(a)). Early- and late-type galaxies show broadly similar distributions of the bulge and disk components in the $\lambda_{\rm R_e}$--$\epsilon$ plane. However, we do detect different kinematic behaviours of bulge components between early- and late-type galaxies, with bulges from late-type galaxies spread toward higher $\lambda_{\rm R_e}$ and $\epsilon$. Accordingly, we find that 63\% and 43\% of bulges from early- and late-type galaxies are classified as slow rotators, suggesting more bulges from late-type galaxies are rotationally supported than those from early-type galaxies. This result is consistent with our finding of a higher mean rotation velocity in late-type bulges (Table~\ref{tab:scatter}).

The separation of the two components in the $\lambda_{\rm R_e}$--$\epsilon$ plane is also reported by Tabor et~al. (2019), based on a similar method but using early-type galaxies from MaNGA IFS data. Considering the similarity in the data, method and sample, it is possible to directly compare Figure~\ref{lambmor}(b) to Figure 6 of Tabor et~al.\ (2019). Both studies found that bulge components have low $\lambda_{\rm R_e}$ and $\epsilon$ indicating they are pressure-supported systems, though we detect a slightly higher fraction of fast-rotating bulges compared to their study. However, we also found a noticeable difference in the $\lambda_{\rm R_e}$--$\epsilon$ distribution of disk components between the two studies: disk components are more elongated (higher $\epsilon$) and rotationally-supported (higher $\lambda_{\rm R_e}$) in this study compared to that of Tabor et~al.\ (2019). The majority of disk components from Tabor et~al.\ (2019) are located between the demarcation line and the theoretical line for an intrinsic ellipticity of 0.4, whereas our disk components are mostly located above the theoretical line for an intrinsic ellipticity of 0.75. 

We found two main reasons for the discrepancy in disk $\lambda_{\rm R_e}$. First, the kinematic map of the disk component from Tabor et~al.\ (2019) does not include central spaxels where the bulge fraction is mostly higher than 0.7. In contrast, most of our disk components from early-type galaxies still have kinematic values in central spaxels (see Figure A1). Therefore, disk $\lambda_{\rm R_e}$ from Tabor et~al.\ (2019) will be biased towards the value in outer bins. This still does not explain the lower $\lambda_{\rm R_e}$ of their disk component, because the $\lambda_{\rm R_e}$ measured excluding central spaxels will generally be higher than that including central spaxels, due to the rapid increase in $\lambda_{\rm R_e}$ within 1\,$R_{\rm e}$ (e.g. Figure~\ref{lambp}). Instead, we believe the reason for the discrepancy arises from the use of binned data. The use of binned data by Tabor et al. (2019) makes their velocity dispersion ($\sigma_{\rm bin}$) an over-estimate relative to individual spaxels, as the velocity gradient $\Delta V$ can be substantial for larger bins and $\sigma_{\rm bin} \sim \sqrt{ (\Delta V)^2+\sigma^2}$. Thus $\sigma_{\rm bin}$ can be a significant over-estimate of $\sigma$ for disk components with large velocity gradients in outer bins with large sizes.

Our $\lambda_{\rm R}$ measurements are not corrected for seeing. Harborne et~al.\ (2019) tested the effect of seeing on the observed $\lambda_{\rm R}$ using mock IFS observations, and reported that in the typical SAMI seeing conditions $\lambda_{\rm R}$ decreases by as much $\sim0.05-0.2$ across the range of galaxy types relative to the value without spatial blurring. Therefore, our $\lambda_{\rm R}$ is thus a lower limit on the intrinsic $\lambda_{\rm R}$ under the seeing condition. However, the predicted amount of the correction in $\lambda_{\rm R}$ suggests that spatial blurring cannot generate the clear separation into two components in the $\lambda_{\rm R}$ distribution.  

Both bulge and disk components share the same potential within galaxies and so all the components are expected to show the same second order velocity moments ($V_{\rm rms} \equiv \sqrt{V_{\rm rot}^2 + \sigma_{\rm e}^2}$). The relative differences ($V_{\rm rms,component}/V_{\rm rms,galaxy}$) are, respectively, +1\% and -3\% for bulge and disk components. Again, this suggests the separation in $\lambda_{\rm R}$ and $V/\sigma$ of two components is less likely to be an artefact of the decomposition. 

Detailed studies show that individual galaxies have complex star-formation and assembly histories, with several stellar-population and kinematic components, corresponding to various star-formation and accretion events (e.g.\ Bond et~al.\ 2010; Zhu et~al.\ 2018; Poci et~al.\ 2019). However, our relatively simple kinematic bulge-disk decomposition conveniently captures the properties of the various components, by grouping together dissipative episodes (e.g.\ star-formation following gas-rich mergers/accretion) and dissipationless episodes (e.g.\ gas-poor mergers/accretion) and provides evidence in support of distinct kinematics produced by these two channels.

Although we find observational evidence for distinct kinematics between bulge and disk components, this has not been confirmed by dynamical modelling. Previous studies reported that the kinematics of fast-rotating early-type galaxies (i.e.\ lenticulars) are well reproduced using a single dynamical model suggesting similar anisotropies in bulge and disk components. Tabor et~al.\ (2019) tested the Jeans Anisotropic Modelling method (JAM; Cappellari 2008) with two components and could not find statistically significant improvement in the model having two distinct kinematics. The significance of our empirical decomposition requires dynamical modelling to be properly understood.

\section{Discussion}
\label{sec:dis}

\subsection{Bulge fraction determines galaxy kinematics}
\label{reconstruction}
The results in Section~\ref{sec:res} indicate that the bulge and disk components have distinct kinematics. Although galaxies show a variation in kinematics according to galaxy types (e.g.\ morphology, B/T), each component shows relatively similar kinematics regardless of galaxy types. Then, can the combination of two distinct components explain the dependence of the galaxy kinematics on galaxy types? The fraction of the bulge (or disk) component well quantifies the types of galaxies, and therefore, can be used to test the importance of the proportion of two components in generating the galaxy kinematics.

\begin{figure}
\centering
\includegraphics[width=\columnwidth]{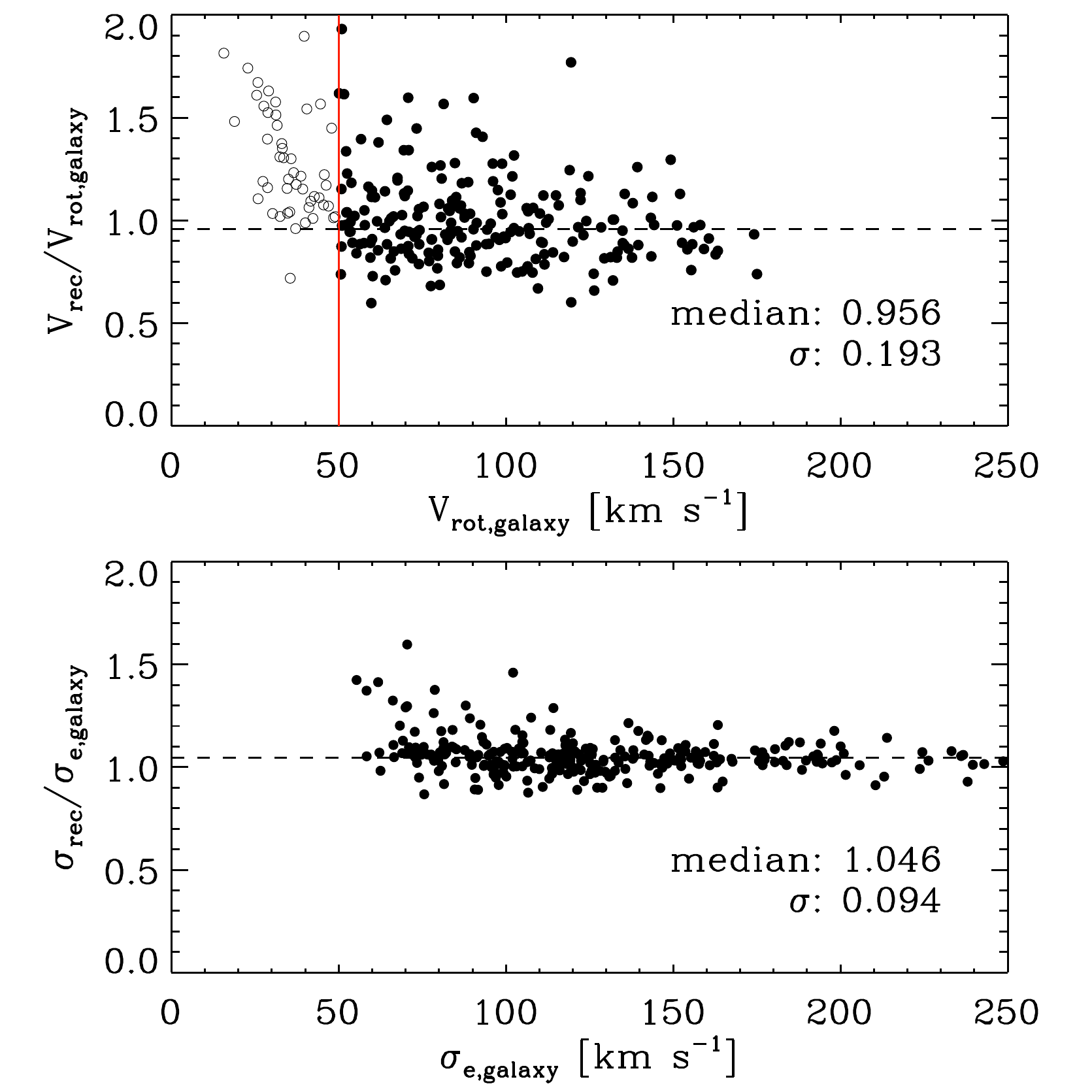}
\caption{A comparison of the reconstructed and observed rotation velocities (top) and velocity dispersions (bottom). The median (dashed line) and standard deviation of the ratio of reconstructed to observed values are shown in each panel. Note that galaxies with $V_{\rm rot,galaxy} < 50$\,km\,s$^{-1}$ (denoted by open circles) are excluded from the statistics for the top panel. The observed and reconstructed quantities are in agreement to within about 5\% in the median in both cases, with 20\% and 10\% uncertainties, respectively, in $V_{\rm rot}$ and $\sigma_{\rm e}$.}
\label{kbt}
\end{figure}

We reconstructed the kinematics of galaxies as the flux-weighted sum of two decomposed kinematics using B/T$_{\rm e}$:
\begin{equation}
\label{eq:vbt}
V_{\rm rec} \equiv B/T_{\rm e} \times V_{\rm rot,bulge} + (1-B/T_{\rm e}) \times V_{\rm rot,disk},
\end{equation}
\begin{equation}
\label{eq:sbt}
\sigma_{\rm rec}^2 \equiv B/T_{\rm e} \times \sigma_{e,bulge}^2 + (1-B/T_{\rm e}) \times \sigma_{e,disk}^2.
\end{equation}
The reconstructed $V_{\rm rec}$ and $\sigma_{\rm rec}$ have similar values to measured quantities (Figure~\ref{kbt}). The $\sigma_{\rm rec}$ agrees well with $\sigma_{e,galaxy}$ with a 10\% uncertainty. The observed and reconstructed velocities are also in agreement within a 20\% uncertainty, and the difference between the $V_{\rm rot,galaxy}$ and $V_{\rm rec}$ is large in galaxies with low levels of rotations (e.g.\ $V_{\rm rot,galaxy} < 50$\,km\,s$^{-1}$). The $V_{\rm rot,galaxy}$ in this study is not measured as a flux-weighted value (Equation~\ref{eq:vel}), unlike the velocity dispersion (Equation~\ref{eq:sig}), which may cause the larger uncertainty in $V_{\rm rec}$. 

Moreover, the measurement error becomes significantly larger when the rotation velocity is small (Figure~\ref{err}), which also produces the large difference in the rotation velocity at low levels. The agreement between the reconstructed and measured kinematics suggests that galaxy kinematics can be analytically described by the combination of two distinct components. The dependence of galaxy kinematics to galaxy types can also be explained by the increasing fraction of the bulge component in earlier morphology. 

One may think that Equations~\ref{eq:vbt} and \ref{eq:sbt} are directly related to the decomposition procedure of pPXF (Section~\ref{sec:ppxf}). However, this analysis is valid for the aperture kinematics which is expected to follow the global potential and where the net rotation velocities of both components are expected to be zero. The $\sigma_{\rm rec}$ is not always comparable to $\sigma_{\rm galaxy}$ in individual spaxels because $\sigma_{\rm galaxy}$ cannot be approximated as a simple sum of two Gaussians especially when two components have different rotation velocities. We have tested this using all the spaxels from 282 galaxies which are available for measuring both bulge and disk kinematics. When the velocity difference of two components is less than 30\,km\,s$^{-1}$ the median $\sigma_{\rm rec}/\sigma_{\rm galaxy}$ is 1.1, and the median $\sigma_{\rm rec}/\sigma_{\rm rot,galaxy}$ continuously decreases as the velocity difference increases. The median values of $\sigma_{\rm rec}/\sigma_{\rm rot,galaxy}$ are, respectively, 0.96 and 0.83 for the velocity differences of 100 and 150\,km\,s$^{-1}$.

\subsection{Intrinsic shape of bulges and disks}
The intrinsic (edge-on) ellipticity ($\epsilon_{int}$) has been estimated using the observed $V/\sigma$ and $\epsilon$ based on the theoretical prediction of intrinsic shape for varying inclination under the assumption of rotating oblate system (Binney 2005). Cappellari (2007) has shown that intrinsic ellipticity correlates with the anisotropy parameter using early-type galaxies from the SAURON IFS survey (de Zeeuw et~al.\ 2002). Cappellari et~al.\ (2016) presented the relationship between the intrinsic and observed ellipticity: 
\begin{equation}
\label{eq:iell}
\epsilon = 1-\sqrt{1+\epsilon_{\rm int} (\epsilon_{\rm int}-2) \sin^2 i}.
\end{equation}
The observed $V/\sigma$ with an inclination $i$ can be corrected to the intrinsic $(V/\sigma)_{\rm int}$ for an edge-on ($i = 90\deg$) view (Binney \& Tremaine 1987):
\begin{equation}
\label{eq:ivsig}
\left(\frac{V}{\sigma}\right) = \left(\frac{V}{\sigma}\right)_{\rm int} \frac{\sin i}{\sqrt{1-\delta \cos^2 i}},
\end{equation}
where $\delta$ is an approximation to the anisotropy parameter $\beta$ such as $\delta \approx \beta = 0.7 \epsilon_{\rm int}$. For each component, we generated a grid with varying $\epsilon_{\rm int}$ and $i$. Then, we calculated corresponding $\epsilon$ and $V/\sigma$ for each $\epsilon_{\rm int}$ and $i$ using Equations~\ref{eq:iell} and \ref{eq:ivsig}. The $\epsilon_{\rm int}$ has been derived from the grid which has the best matched $\epsilon$ and $V/\sigma$ to the observation. Note that $V/\sigma$ for the bulge and disk components are measured using $R_{\rm e}$, PA, and $\epsilon$ of each component in this section, whereas all kinematic measurements in the other sections except for this section and Figure~\ref{lambmor} used the structural parameters from galaxies. Measuring $V/\sigma$ at the effective radius of each component gives more galaxies with kinematics for both components: 664 and 633 are available for measuring bulge and disk components (respectively), and 468 galaxies have kinematics available for both components. On the other hand, the above method is not valid for slowly rotating bulges (e.g.\ below the green line in Figure~\ref{lambmor}) because they are not oblate rotators. In that instance, we adopt $\epsilon_{int}=\epsilon$ as our best estimate of the intrinsic ellipticity (although it is strictly a lower limit).  

\begin{figure}
\centering
\includegraphics[width=\columnwidth]{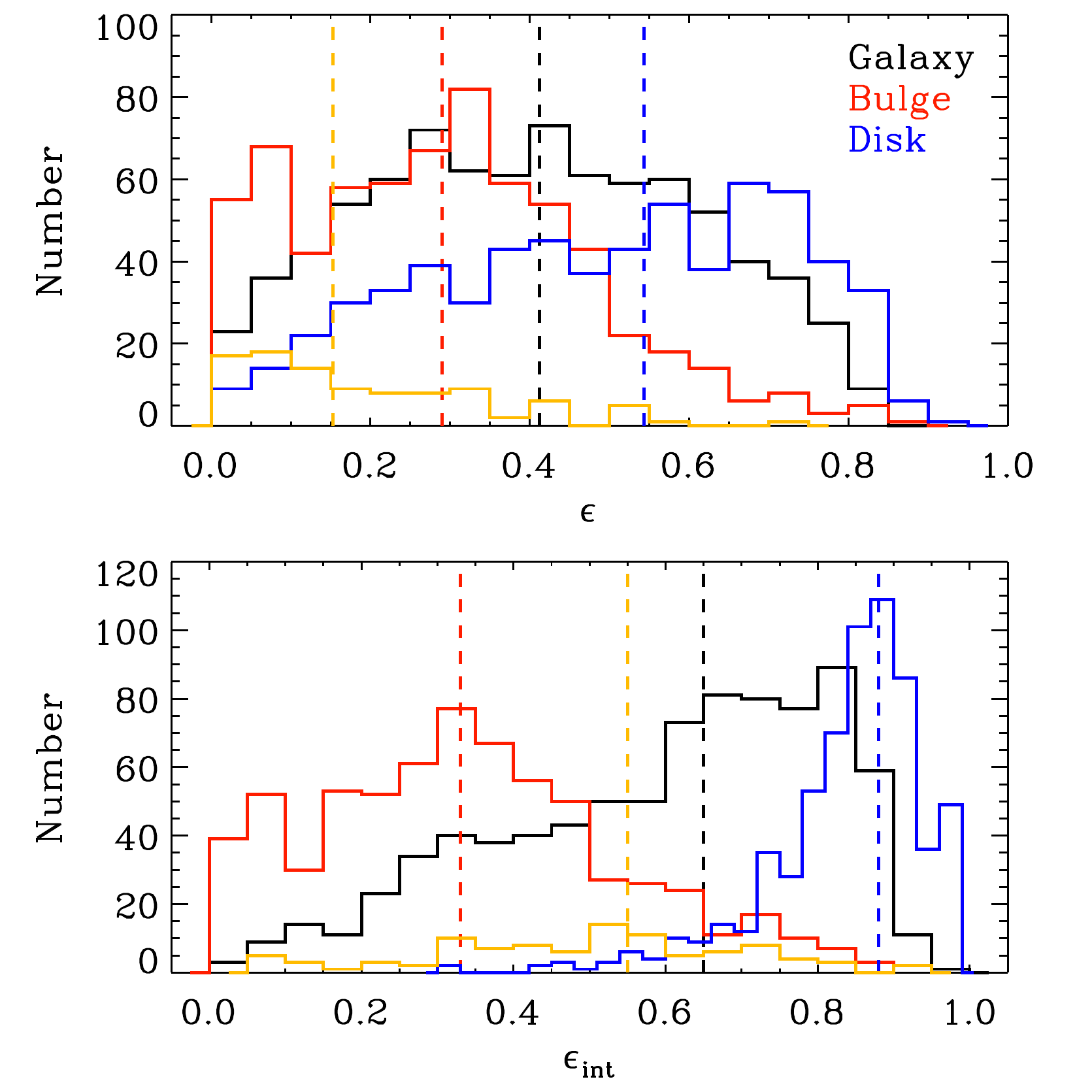}
\caption{The distributions of apparent ellipticity ($\epsilon$; top) and intrinsic ellipticity ($\epsilon_{\rm int}$; bottom). Black, red and blue histograms show, respectively, galaxies and their bulge and disk components; the dashed lines show the median of each distribution. The yellow histogram shows a subsample of bulge components that have higher $V/\sigma$ than that expected for oblate rotators in edge-on view (e.g.\ the green line in Figure~\ref{lambmor}), and their $\epsilon_{\rm int}$ is estimated as described in the text. The change in going from $\epsilon$ to $\epsilon_{\rm int}$ is substantial for galaxies and their disk components. }
\label{ell}
\end{figure}

We present the distribution of observed and intrinsic ellipticity in Figure~\ref{ell}. Our sample galaxies have a broad range of $\epsilon$. The distributions of $\epsilon$ for the bulge and disk components are slightly skewed towards low and high values in $\epsilon$, respectively, but also show a wide range. The distribution of $\epsilon_{\rm int}$ differs from that of $\epsilon$, especially for galaxies and disk components. The correction for the inclination is large in the ellipticity of galaxies, and the distribution of $\epsilon_{\rm int}$ is highly skewed towards high values (see also van de Sande et~al.\ 2018). In general, disk components show a high $\epsilon_{\rm int}$ with a median value of 0.88, confirming that disk components are intrinsically flattened systems and have a small variation in flattening. The median $\epsilon_{\rm int}$ of bulge components (0.34) shows a similar value to that of slow rotators (0.37) presented in Weijmans et~al.\ (2014). We do not find a difference between $\epsilon$ and $\epsilon_{\rm int}$ in bulge components because many bulges have a lower $V/\sigma$ than the intrinsic one expected for oblate rotators at a given $\epsilon$, and $\epsilon_{\rm int}$ and $\epsilon$ have the same value assuming $i = 90^{\deg}$. The distribution of $\epsilon_{\rm int}$ also supports that the bulge and disk components have distinct dynamical and structural features.

We also present the distribution of ellipticity for a subsample of bulge components whose observed $V/\sigma$ is above the theoretical expectation for the edge-on view of oblate rotators (e.g.\ the green line in Figure~\ref{lambmor}), and their $\epsilon_{\rm int}$ is estimated as following the above method. The $\epsilon_{\rm int}$ of the subsample is spread over a wide range, generating a tail in the distribution of bulge $\epsilon_{\rm int}$ toward higher value. We still find a distinct $\epsilon_{\rm int}$ distribution for the subsample compared to that from disk components.

\subsection{Uncertainty in $f_{\rm bulge}$}
The method described in Section~\ref{sec:decomp} depends strongly on the relative weight of the two components. Our kinematic measurements therefore reflect the uncertainty in photometric decomposition. Estimating realistic uncertainties in $f_{\rm bulge}$ is difficult due to various factors (e.g.\ inclination, bar/spiral/ring structures, and mergers). We examined the impact of bar structures in our results because there is the possibility of significantly over-estimating bulge light for strongly-barred galaxies. We visually inspected the 295 galaxies in the late-type sample and identified 56 (strongly) barred galaxies, of which 16 galaxies are available for measuring bulge kinematics at $R_{\rm e}$ for the galaxy as a whole. Barred galaxies, therefore, are rare in our bulge kinematics. We examined the location of barred galaxies in the scaling relations and the $\lambda_{\rm R_e}$--$\epsilon$ plane and found that barred galaxies show similar $V_{\rm rot}$, $\sigma_{\rm e}$, and $\lambda_{\rm R_e}$ compared to the rest of the sample. Therefore our results are expected to be stable, considering the small number of barred galaxies in our sample and the small impact of bar structures in $V_{\rm rot}$, $\sigma_{\rm e}$, and $\lambda_{\rm R_e}$.

We also estimated uncertainties in $V_{\rm rot}$ and $\sigma_{\rm e}$ assuming conservative uncertainties in $f_{\rm bulge}$ ($\pm10\%$). For 75 high-S/N SAMI galaxies that are available for measuring both bulge and disk kinematics, we applied the same method described in Section~\ref{sec:ppxf} with $f_{\rm bulge}\pm0.1$. When we forced $f_{\rm bulge}$ to be lower than the original value by 0.1, we found 15\% and 8\% decreases, respectively, in bulge and disk $V_{\rm rot}$, together with 5\% and 9\% increases, respectively, in bulge and disk $\sigma_{\rm e}$. When we forced $f_{\rm bulge}$ to be higher than the original value by 0.1, we found 15\% and 6\% increases, respectively, in bulge and disk $V_{\rm rot}$, together with 5\% and 10\% decreases, respectively, in bulge and disk $\sigma_{\rm e}$. This indicates conservative uncertainties in our $V_{\rm rot}$ and $\sigma_{\rm e}$ measurements, when changing the bulge fraction. This suggests the uncertainty in $f_{\rm bulge}$ does not tend to make the two components more kinematically distinct. Nevertheless, the uncertainly in $f_{\rm bulge}$ is expected to add an extra scatter in the kinematics of two components, which can be reflected in the scatter of the scaling relations.

\section{Conclusion and Summary}
\label{sec:con}
We investigate the kinematics of 826 galaxies from the SAMI Galaxy Survey which have been selected based on the stellar mass, size, and S/N (Section~\ref{sec:sam}). Our sample includes 295 late-type galaxies, of which 116 have bulge S\'ersic n lower than 1.5. Only 23 of the 116 late-types are available to measure bulge kinematics, so that our results mainly describe the kinematic behaviour of `classical' bulges. We spectroscopically decomposed bulge and disk kinematics using photometrically-defined flux weights. The rotation velocity and velocity dispersion of two components have been simultaneously estimated for each spaxel using pPXF with a new subroutine to overcome the degeneracy in $\chi^2$ and trace a physically plausible solution.

The Faber-Jackson and Tully-Fisher relations indicate that $V_{\rm rot}$ and $\sigma_{\rm e}$ of both bulge and disk components scale with the stellar mass. As expected, disk components follow the Tully-Fisher relation well for the low-B/T galaxies (B/T~$< 0.2$), and bulge components have a similar Faber-Jackson relation to the high-B/T galaxies (B/T~$>0.8$). Therefore, two scaling relations stem from either the rotation (disk) or dispersion (bulge)-dominated components of galaxies. We also detected a tight correlation between the stellar mass and the velocity dispersion of the disk component. 

The sample galaxies with various bulge fractions show a large scatter in the scaling relations. Moreover, the magnitudes of residuals from the scaling relations correlate with galaxy types (e.g.\ morphology and B/T). Galaxies with a higher B/T are more scattered from the Tully-Fisher relation, and galaxies with a low B/T mainly cause the scatter in the Faber-Jackson relation. On the other hand, the kinematics of the bulge and disk components scale well with the stellar mass and do not show a residual trend with galaxy types. Therefore, the scaling relations do not apply to a particular type of galaxy, but rather describe a general scaling of either the rotation- or dispersion-dominated component of galaxies. The dominance of one component in the low-B/T or high-B/T galaxies promotes following either the Tully-Fisher relation or the Faber-Jackson relation. 

According to our results, bulge and disk components are kinematically distinct, and the weights of two components seem to take a major role in determining the kinematics of galaxies. It has been reported that the spin parameter $\lambda_{\rm R}$ correlates with galaxy properties, such as mass and age (e.g.\ van de Sande et~al.\ 2018). We also found a negative correlation between the $\lambda_{\rm R}$ of galaxies and the bulge fraction. Bulge and disk components exhibit well-separated distributions in $\lambda_{\rm R_e}$ and $V/\sigma$ suggesting that bulges are dispersion-dominated, and disks are supported by rotation. Moreover, the $\lambda_{\rm R}$ of two components does not show a clear connection to the B/T. Our results show that the diverse kinematic behaviour of galaxies can be explained (at least to first order) as relative contributions of disk and bulge components. We were also able to reproduce the kinematics of galaxies based on the weights and decomposed kinematics of the two components. Our results support the interpretation of the kinematics of galaxies as the combination of two distinct kinematic components, with the overall properties mainly determined by the relative weights of these components.

\section*{Acknowledgements}
This research was supported by the Australian Research Council Centre of Excellence for All Sky Astrophysics in 3 Dimensions (ASTRO 3D), through project number CE170100013. The SAMI Galaxy Survey is based on observations made at the Anglo-Australian Telescope. The Sydney-AAO Multi-object Integral field spectrograph (SAMI) was developed jointly by the University of Sydney and the Australian Astronomical Observatory. The SAMI input catalogue is based on data taken from the Sloan Digital Sky Survey, the Galaxy And Mass Assembly (GAMA) Survey and the VLT Survey Telescope (VST) ATLAS Survey. The SAMI Galaxy Survey is supported by the Australian Research Council Centre of Excellence for All Sky Astrophysics in 3 Dimensions (ASTRO 3D), through project number CE170100013, the Australian Research Council Centre of Excellence for All-sky Astrophysics (CAASTRO), through project number CE110001020, and other participating institutions. The SAMI Galaxy Survey website is sami-survey.org. This research used Penalized Pixel-Fitting method by Cappellari \& Emsellem (2004) as upgraded in Cappellari (2017). We are grateful to the anonymous referee for constructive comments and suggestions, especially on the different kinematic behaviour of late-type bulges (Section~\ref{sec:lamb}). SO thanks DL for the consistent support. LC is the recipient of an Australian Research Council Future Fellowship (FT180100066) funded by the Australian Government. JJB acknowledges support of an Australian Research Council Future Fellowship (FT180100231). FDE acknowledges funding through the H2020 ERC Consolidator Grant 683184. JBH is supported by an ARC Laureate Fellowship (FL140100278) that funds JvdS and an ARC Federation Fellowship that funded the SAMI prototype. MSO acknowledges the funding support from the Australian Research Council through a Future Fellowship (FT140100255). NS acknowledges support of an Australian Research Council Discovery Early Career Research Award (DE190100375) funded by the Australian Government and a University of Sydney Postdoctoral Research Fellowship.

\appendix
\section{Sample Kinematic maps}
\label{app:kin}
\begin{figure*}
\centering
\includegraphics[width=0.98\textwidth]{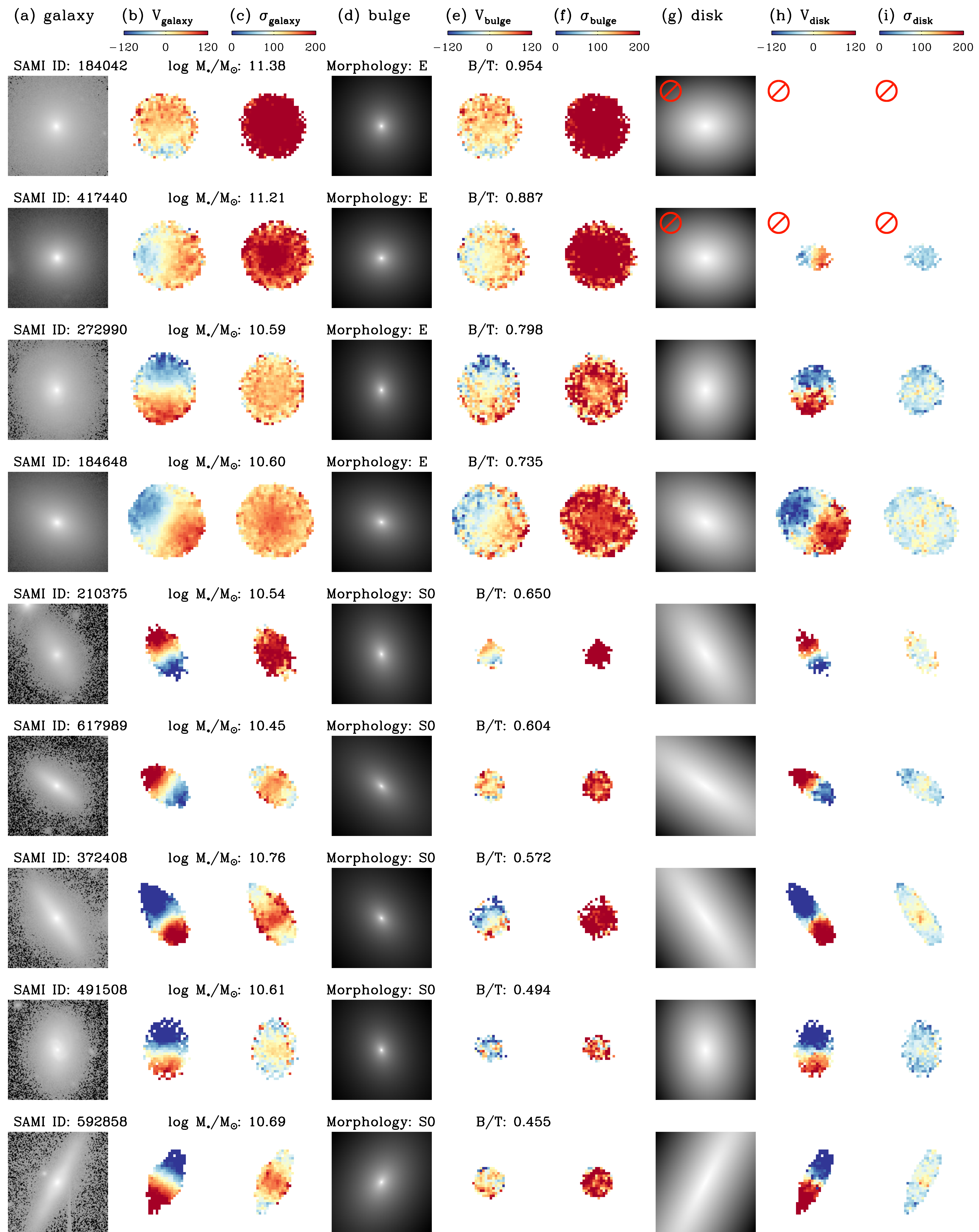}
\caption{Sample kinematic maps of early-type galaxies: (a) the $r$-band image; (b) galaxy $V_{\rm rot}$; (c) galaxy $\sigma_{\rm e}$; (d) bulge model; (e) bulge $V_{\rm rot}$; (f) bulge $\sigma_{\rm e}$; (g) disk model; (h) disk $V_{\rm rot}$; and (i) disk $\sigma_{\rm e}$. Images and models are presented in log scale. All velocities are presented in a range between -120 and 120\,km\,s$^{-1}$, and all velocity dispersions are shown in the range between 0 and 200\,km\,s$^{-1}$. The galaxies are ordered by B/T. A `prohibited' symbol ($\oslash$) is shown at top-left when the component is not included in our sample.}
\label{app:kin1}
\end{figure*}

\begin{figure*}
\centering
\includegraphics[width=0.98\textwidth]{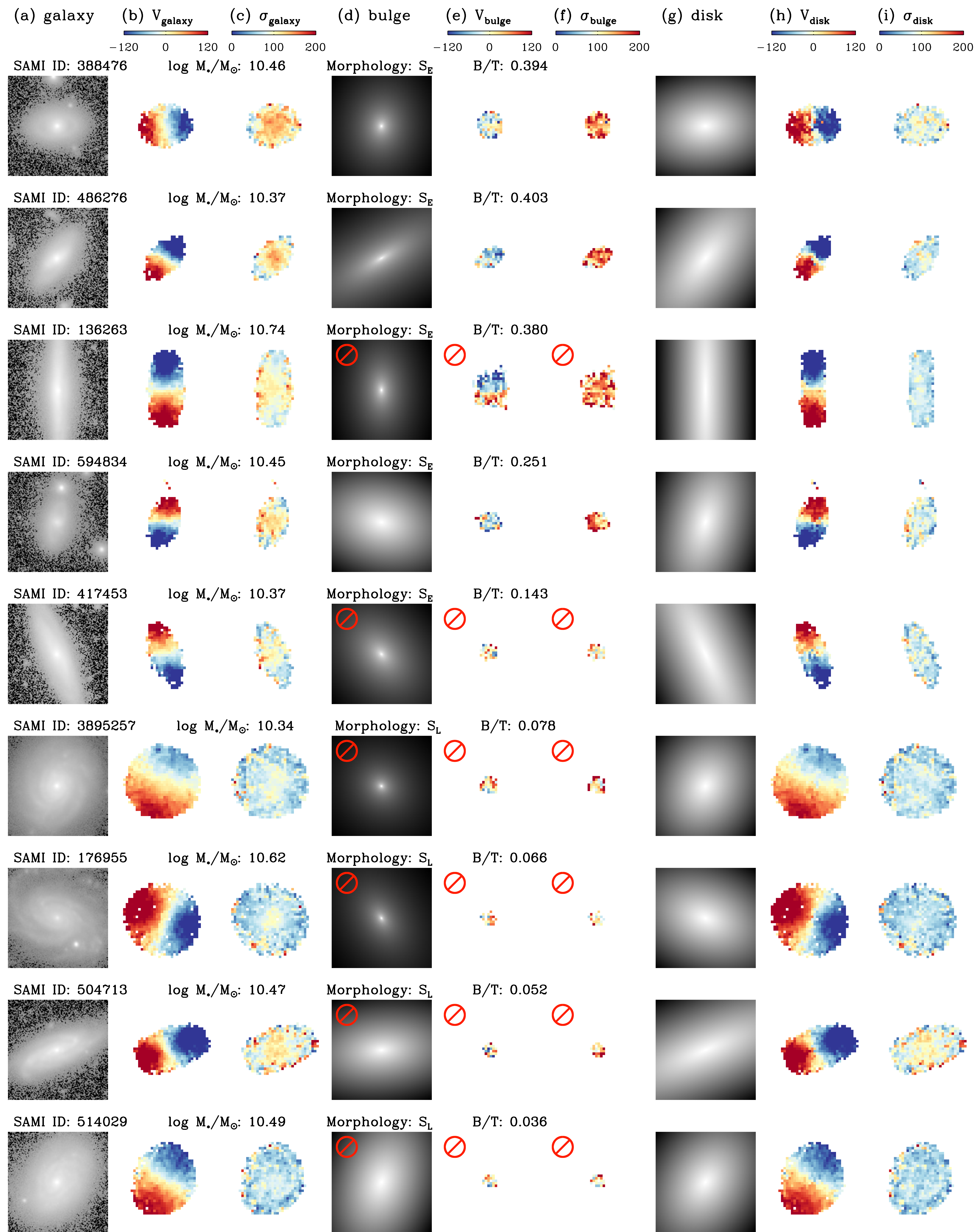}
\caption{Sample kinematic maps of late-type galaxies. Details are the same as described in Figure~\ref{app:kin1}.}
\label{app:kin2}
\end{figure*}

In Figures~\ref{app:kin1} and \ref{app:kin2}, we present sample kinematic maps for various galaxy types. 

\section{The distribution of $\chi^2$ }
\label{app:chi}
\begin{figure}
      \centering
       \includegraphics[width=\columnwidth]{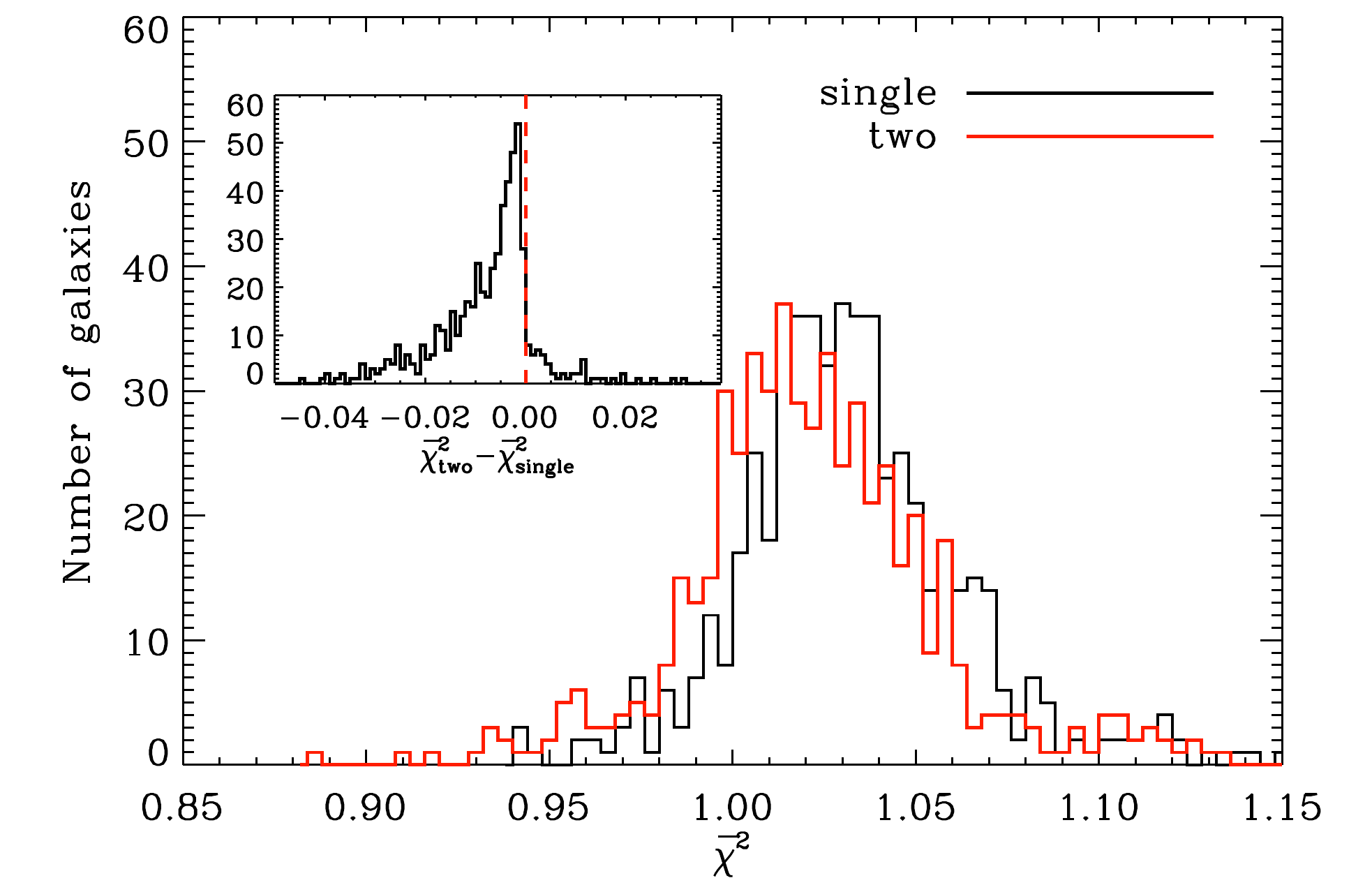}
       \caption{The distribution of the mean reduced $\chi^2$ ($\overline{\chi^2}$) from single- and two-component pPXF fitting. }
       \label{chi}
     \end{figure}

We have obtained reduced $\chi^2$ from pPXF for each spaxel and calculated the mean reduced $\chi^2$ ($\overline{\chi^2}$) within 1\,$R_{\rm e}$ using spaxels whose S/N is greater than 3\,\AA\ to get the representative value for each galaxy. In Figure~\ref{chi}, we present the distribution of $\overline{\chi^2}$ fitting single- ($\overline{\chi^2}_{single}$) and two-component ($\overline{\chi^2}_{two}$) models using pPXF. The $\chi^2$ values from both single- and two-component fits are nearly 1 because the input noise has been normalised based on the residual (see Section~\ref{sec:ppxf}). Note that the normalisation factor has been decided for each spectrum and is the same for both single and two components fit. Although the difference is marginal (partially) due to the normalisation, the $\overline{\chi^2}$ from the pPXF fitting using two components tends to be lower than that using a single component.

 \section{Error estimation from noise shuffling}
 \label{app:err} 
 The SAMI survey team has so far estimated the error for stellar kinematics by simulating 150 mock cubes with shuffled noise for each galaxy (van de Sande et~al.\ 2017b). This method gives more realistic errors than using errors from pPXF, however, running pPXF for decomposing two kinematics requires roughly ten times longer computing time, which discourages us from using the noise shuffling method for estimating errors. Instead, we have applied the method for 15 galaxies with various stellar masses and S/N to examine the bias on the errors from pPXF using the Monte Carlo simulation. We generated hundreds of mock cubes for each galaxy by shuffling residuals between the best fit model and the observed spectra. For each spaxel, the noise has been randomly shuffled within a $\sim$400\,\AA\ window to ensure that the noise signature in mock cubes is comparable to that of the original spectra over the wavelength. Then, we ran pPXF and calculated the standard deviation for the kinematics of hundred mock cubes.  
 
\begin{figure}
\centering
\includegraphics[width=\columnwidth]{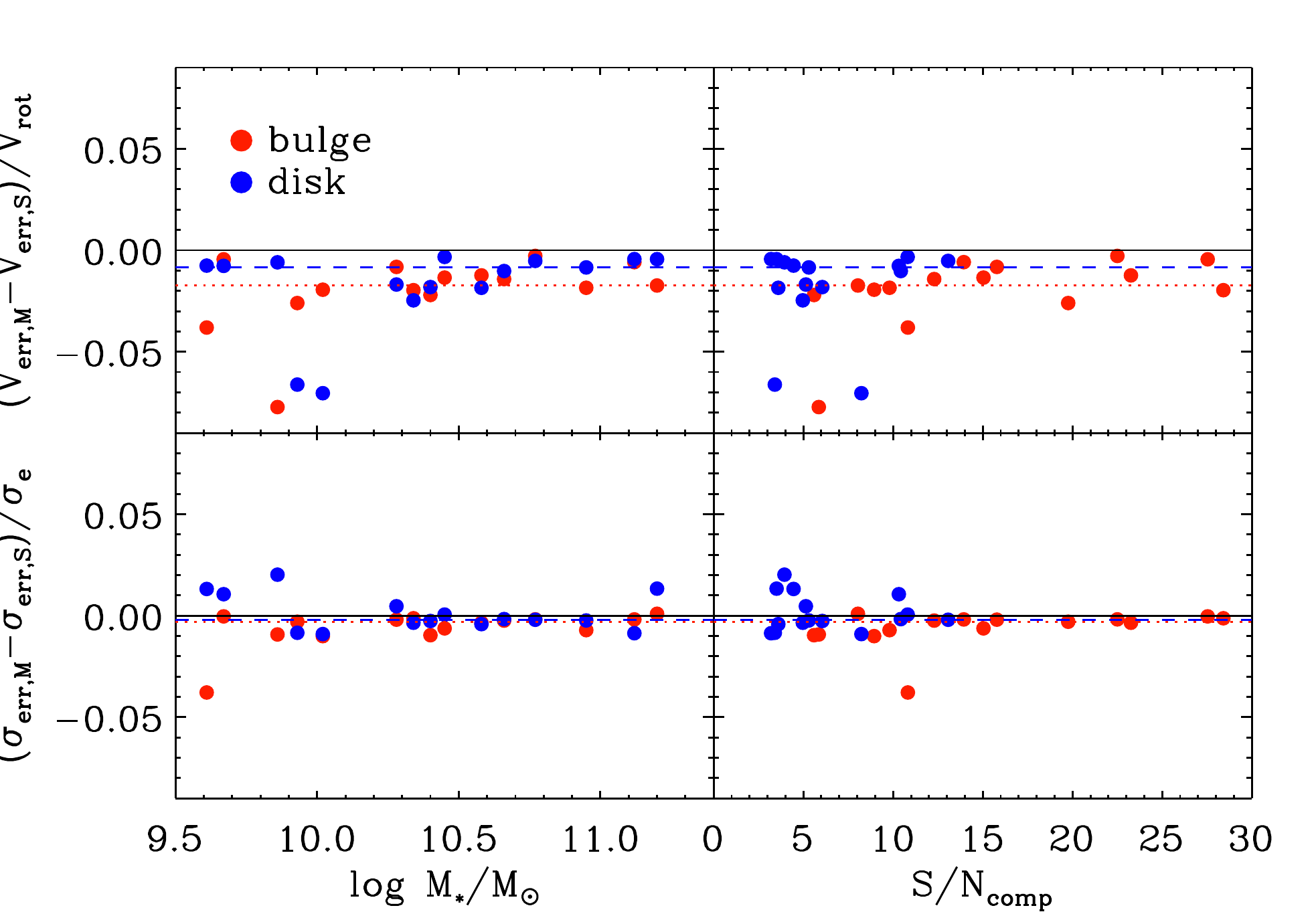} 
\caption{The relative difference in errors of the kinematics between Monte Carlo ($V_{\rm err,M}$ and $\sigma_{\rm err,M}$) and noise shuffling ($V_{\rm err,S}$ and $\sigma_{\rm err,S}$) methods. Dashed and dotted lines show, respectively, the median difference in the errors for the bulge and disk kinematics.}
\label{shuffling}
\end{figure}

 In Figure~\ref{shuffling}, we compare the relative errors in the kinematics measured within 1\,$R_{\rm e}$ from the Monte Carlo (Section~\ref{sec:mkin}) and the noise shuffling methods. The Monte Carlo method tends to underestimate the error in rotation velocity compared to the noise shuffling method by 2\% for the bulge and 1\% for the disk components. Also, we found some outliers when stellar mass or S/N is low. Both methods yield similar errors in velocity dispersion. Therefore, we conclude that the error estimated from the Monte Carlo simulation used in this study is comparable to that based on noise shuffling. 
 
\section{Annular Kinematics}
\label{app:beam}
\begin{figure}
\centering
\includegraphics[width=\columnwidth]{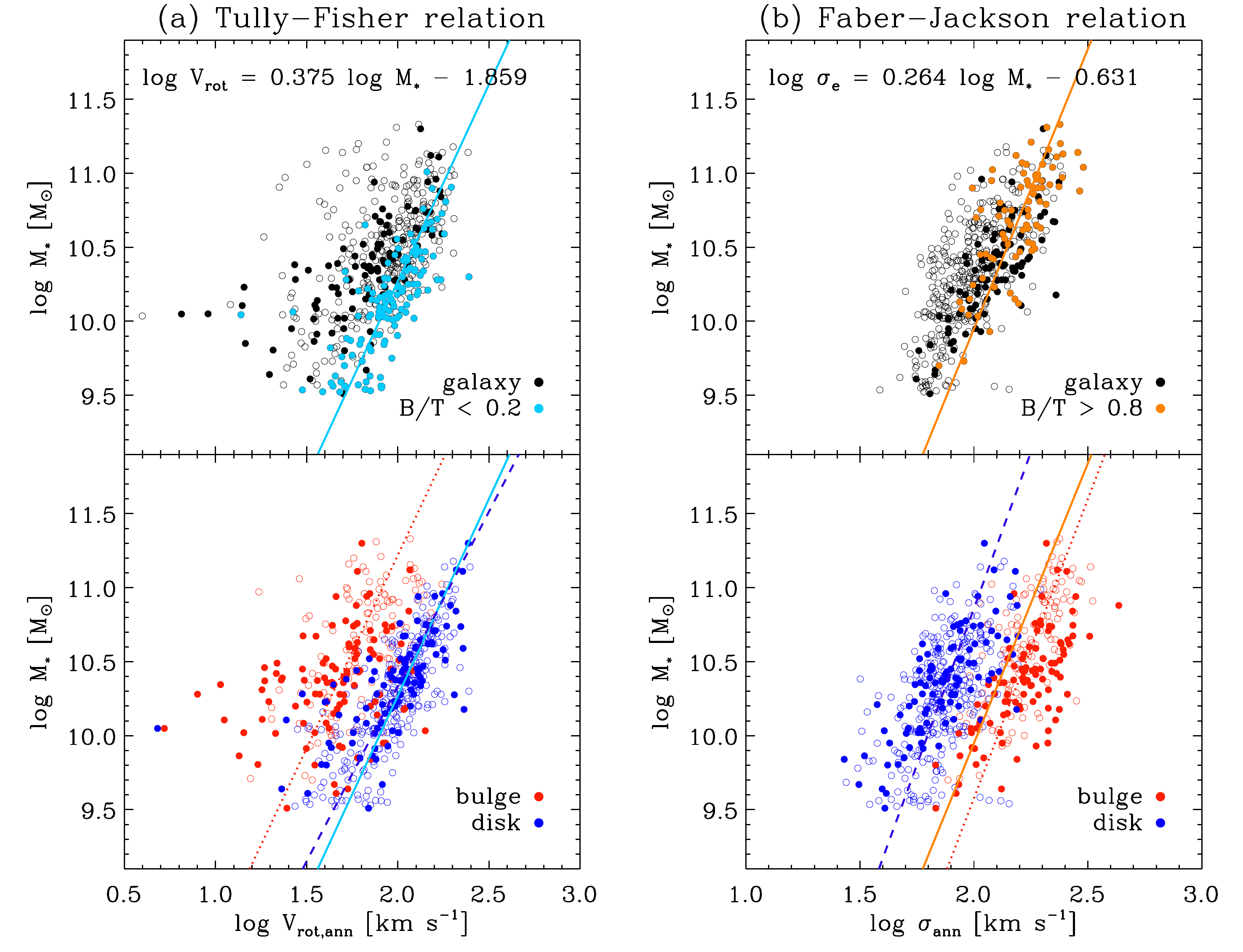}
\caption{The Tully-Fisher and Faber-Jackson relations using annular kinematics ($V_{\rm rot,ann}$ and $\sigma_{ann}$) measured within $0.85<R_{\rm e}<1.15$. Details are the same as described in Figure~\ref{scaling}.}
\label{app:ann1}
\end{figure}

 The impact of beam smearing on the measured velocity dispersion is maximised when the relative size of the galaxy is small compared to the PSF and when the velocity gradient is large (e.g.\ Johnson et~al.\ 2017; Harborne et~al.\ 2019). The velocity dispersion in this study has been cumulatively measured within 1\,$R_{\rm e}$, and therefore, there is a possibility that our measurement of velocity dispersion has been affected by beam smearing. The impact of beam smearing rapidly increase toward the centre of galaxies, and the velocity dispersion measured within an annulus excluding the centre of galaxies can reduce the impact of beam smearing. 
 
\begin{figure}
\centering
\includegraphics[width=\columnwidth]{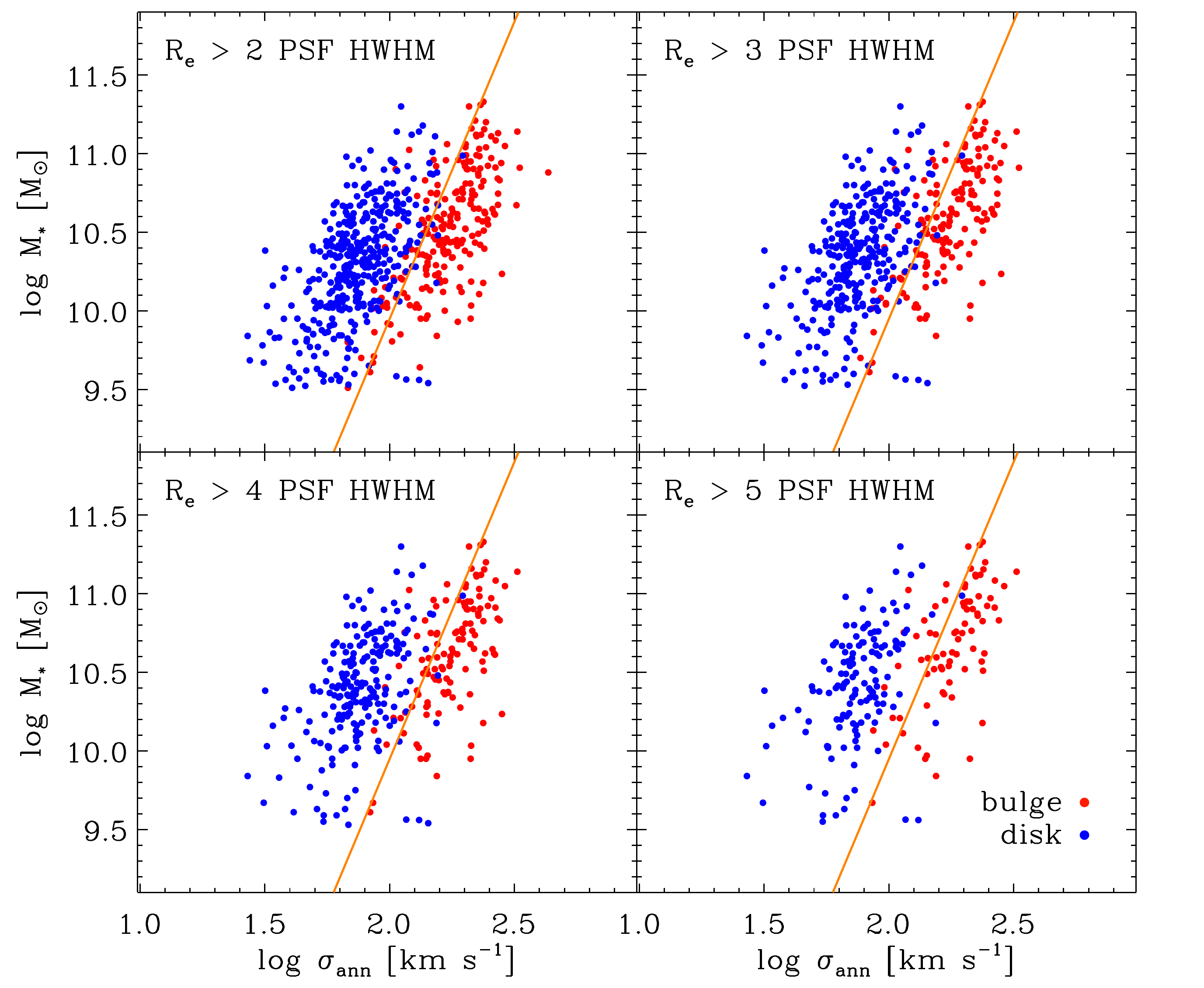}
\caption{The Faber-Jackson relation of the bulge (red) and disk (blue) components using annular kinematics for various galaxy sizes. Details are the same as described in Figure~\ref{scaling}(b).}
\label{app:ann2}
\end{figure}

\begin{figure}
\centering
\includegraphics[width=\columnwidth]{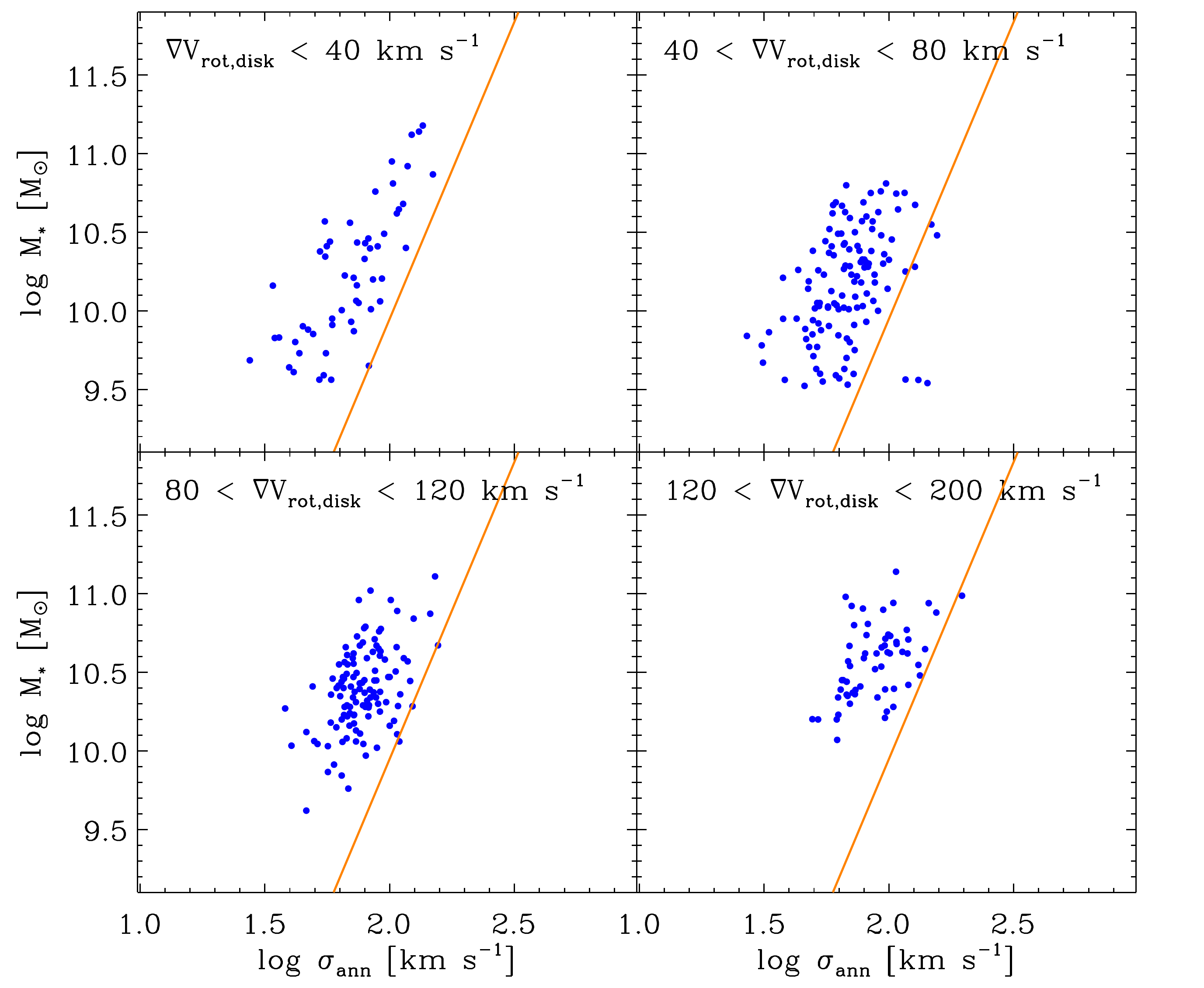}
\caption{The Faber-Jackson relation of the disk components using annular kinematics for different gradients in $V_{\rm rot,disk}$ ($\nabla V_{\rm rot,disk}$) within 1\,$R_{\rm e}$. Details are the same as described in Figure~\ref{scaling}(b).}
\label{app:ann3}
\end{figure}

 We present the Tully-Fisher and Faber-Jackson relations using annular kinematics ($V_{\rm rot,ann}$ and $\sigma_{ann}$) measured within $0.85 < R_{\rm e} < 1.15$ in Figure~\ref{app:ann1}. Although the slopes of the scaling relations are slightly changed, the overall trends are the same to the scaling relations based on the cumulatively measured kinematics (Figure~\ref{scaling}). Moreover, we still find a correlation between the stellar mass and the velocity dispersion in the disk components even using annular kinematics. In Figure~\ref{app:ann2}, we present the Faber-Jackson relations of the bulge and disk components for different apparent galaxy sizes. For both the bulge and disk components, the Faber-Jackson relation still holds in galaxies larger than four times the PSF.  Also, we found the Faber-Jackson relation holds for the disk components of galaxies with a small velocity gradient within 1\,$R_{\rm e}$ ($\nabla V_{\rm rot,disk}$) in Figure~\ref{app:ann3}. Therefore, we conclude that the Faber-Jackson relation in the disk components is not due to beam smearing. 

\label{lastpage}
\end{document}